\documentclass[preprintnumbers,article,amsmath,amssymb,floatfix,10pt,prd,onecolumn,
superscriptaddress,nofootinbib]{revtex4}
\usepackage{bbm}
\usepackage{amsfonts}
\usepackage{mathrsfs}
\usepackage{latexsym}

\usepackage{epsfig}
\usepackage{placeins}
\usepackage{float}
\usepackage{graphicx}
\usepackage{amssymb}
\usepackage{amsmath}
\usepackage{dcolumn}
\usepackage{bm}
\usepackage{color}
\usepackage{comment}
\usepackage{xcolor}

\begin{document}

\title{\bf Stellar Structures in $f(\mathcal{G})$ Gravity with Tolman-Kuchowicz Spacetime}

\author{M. Farasat Shamir}
\email{farasat.shamir@nu.edu.pk}
\author{Tayyaba Naz}
\email{tayyaba.naz@nu.edu.pk}
\affiliation{National University of Computer and
Emerging Sciences,\\ Lahore Campus, Pakistan}
\begin{abstract}
This paper is devoted to explore some relativistic configurations of stellar objects for static
spherically symmetric structures in the context of modified $f(\mathcal{G})$ gravity, by exploiting the Tolman-Kuchowicz spacetime \cite{Tolman, Kuch}.
We develop the equations of motion for spherically symmetric spacetime in the presence of anisotropic matter distribution by considering the physically valid expressions of the metric potentials, $\nu=Br^2+2lnC$ and $\lambda=ln(1 +
ar^2+br^4)$. To attain the values of the unknown constants we consider the observational data of $Cen~ X-3$, $EXO  ~1785-248$ and $LMC~ X-4$ star models.
Further, by using evaluated form of the solutions we provide many aspects which are described by the physical status like effective energy density, components of radial and transverse pressure, energy conditions, stability against equilibrium of the forces, speed of sound,  mass-radius relation, surface redshift, compactness parameter, adiabatic index and anisotropic measurement. It is observed that all these features follow physically accepted patterns and the resulting outcome is in the experimental range which depicts the viability of our presented $f(\mathcal{G})$ gravity models.
\end{abstract}

\maketitle

\section{Introduction}
The spatial behavior of accelerated expansion of universe has captured much attention in the modern generation of cosmology and astrophysics
\cite{1, 2}. Recent evolutions in this era of cosmology have exposed new concepts to acquaint the critical and observational innovations for this accelerated expansion of universe. Different observations may provide direct evidence related to the accelerated expansion as a consequence of high red-shift supernova experiments \cite{3}, while large scale structures \cite{Teg} and cosmic microwave background fluctuations \cite{Spergel} yield indirect evidences.
This accelerating expansion of universe is caused due to a mysterious force named as dark energy which retains strong negative pressure. Moreover, it is considered that the mysterious dark energy comprises almost 68\% of the total energy of the universe. Thus, in order to analyze the phenomenon  of acceleration expansion, we need some modifications to the classical theory. Such type of problems lead into the search of modified or extended theory of gravity that might succeed in defining the situations in which the general theory of relativity (GR) given unsatisfactory results. As an alternative to GR, various gravitational modified theories have been proposed in the recent years.
Some of which are $f(R)$, $f(R,T)$, $f(\mathcal{G})$, $f(R,\mathcal{G})$ and $f( \mathcal{G}, T)$ theories of gravity which have been developed by the composition of curvature scalars, topological invariants along their derivatives. To describe the late time acceleration and dark energy issues, the modifications of GR seem seductive. Furthermore, different cosmological approaches and thoughts provided by these theories are helpful to reveal the secrets behind the phenomenon of accelerated expansion of universe \cite{Capozziello}.
In past few years, Einstein's theory of relativity has been
amended  by many scientists. In these valuable modifications one of the most simplest and well known modified theory \cite{Buch} is $f(R)$  attained by supplanting the term Ricci scalar $R$ with an arbitrary function $f(R)$. These alternative theories of gravity paly a vital role to understand the mysterious nature of the universe, which is responsible for the accelerated  expansion of universe \cite{Odintsov1, Odintsov2}.
Another remarkable theory which has attained eminence in the last few years is dubbed with the name, is modified Gauss-Bonnet gravity, also known as $f(\mathcal{G})$ gravity \cite{Noj1}-\cite{Cog2}. This modified theory of gravity was established by modifying the Einstein Hilbert action, by replacing term $R$ with $f(R, \mathcal{G})$. In the scenario of large expansion of universe, the additional Gauss-Bonnet term has resolved the deficiencies of $f(R)$ theory of gravity \cite{Noj1}-\cite{Felice2}. It is considered that $f(\mathcal{G})$ gravity is the simplest form of $f(R, \mathcal{G})$ theory of gravity, which is extensively addressed and assumed very supportive to reconstruct any form of cosmological solutions. Here $f(\mathcal{G})$ being the generic function of Gauss-Bonnet invariant term.
An interesting significance of $f(\mathcal{G})$ modified
gravity \cite{Chiba} is that it may avoid the ghost contributions and supports in regularizing the gravitational action due to
the Gauss-Bonnet invariant quantity. Further to explore various cosmic issues as an alternative to dark energy \cite{Santos1}, the modified $f(\mathcal{G})$ gravity provide an influential platform for this purpose.
Mak and Harko \cite{Mak} investigated exact solutions of Einstein field equations for some standard models with anisotropic background. An analytical formulation for the solutions of field equations with
anisotropic matter source was constructed by Chaisi and Maharaj \cite{Chaisi}. Rahaman et al. \cite{Raha}
extended the technique of Krori and Barua solution to the system of strange star with
MIT bag model. Kalm  et al. \cite{Kalm1, Kalm2} also studied the compact stellar objects by assuming anisotropic source matter under the Krori and Barua metric. The possibility for the existence of higher dimensional compact star
was explored by Bhar et al. \cite{Bhar1}.
The stability analysis and fundamental formation of anisotropic compact stars were analyzed by Zubair  et al. \cite{Zubair1} in $f(R, T)$ theory of gravity.
Further, the investigations on the charged anisotropic solutions for the compact objects were formulated by Maurya et al. \cite{S.K}. Recently, Ilyas \cite{56} discussed compact structures in modified Gauss-Bonnet gravity in the presence of charge.
\\The relativistic massive objects known as compact stars can
be expressed by GR as well as the gravitational extended theories of gravity \cite{Abbas1}-\cite{Cam}.
These compact stars have strong gravitational force due to its very small
size and immensely massive structure. In astrophysics, the study of compact stars have gained much attention. In recent decades it has been considered a vigorous subject of research due to their fascinating physiognomies and structures. Some exact solutions in the result of a collapsing star
with anisotropic stress and heat flux were explored by Goswami et al. in  $f(R)$ gravity \cite{Goswami}.
The equilibrium condition of compact stars is found by Abbas et al. \cite{Abbas4}, further they also analyzed the physical features in the framework of $f(\mathcal{G})$ gravity. The modified theories of gravity have great contribution in studying and examining the nature of compact stellar structures and matter at high densities \cite{Ast2}-\cite{Ast4}.
\\The aim of this paper is to analyze the appearance
of $f(\mathcal{G})$ gravity in modeling of realistic configurations of
compact stellar objects in the presence of Tolman-Kuchowicz spacetime. In particular, we  extend the idea of Jasim et al. \cite{Jasim} in modified Gauss-Bonnet gravity and examine the stability and physical features of compact stars  $Cen~ X-3,~ EXO~ 1785-248$ and $LMC~ X-4$. We investigate the various structural
properties  by choosing the specific $f(\mathcal{G})$ gravity models in the background of anisotropic matter source like evolution of effective energy density and components of radial and tangential pressure, the Tolman-Oppenheimer Volkoff (TOV)
equation, mass radius relation, compactness parameter, surface redshift, the stability as well
as the different energy bounds, for different experimental
data of compact stellar structures.
\\The layout of this paper is organized as follows:
Section $\textbf{2}$ consists of the mathematical formulation of $f(\mathcal{G})$ gravity in the context of anisotropic matter distributions.
Some of viable $f(\mathcal{G})$ gravity models and boundary conditions are demonstrated in section $\textbf{3}$.
In section $\textbf{4}$, we calculate the values of the unknown constant for the chosen values of our model parameters by matching the interior metric to
Schwarzschild's exterior metric. Section $\textbf{5}$ is devoted to scrutinize some physical attributes and also check the viability of different familiar compact stars via graphical analysis. Last section is based on the conclusive remarks.

\section{Equation of Motion for Relativistic Sphere in $f(\mathcal{G})$ Gravity}

To study the stellar configurations of compact stars in modified Gauss-Bonnet gravity, we consider the
most general action for $f(\mathcal{G})$ gravity as follows \cite{Noj1}
\begin{equation}\label{5}
    S = \int d^4x \sqrt{-g} \Bigg[\frac{R}{2\kappa^2}+f(\mathcal{G})\Bigg] + S_m,
\end{equation}
where $R$ is Ricci scalar, $\kappa^2 = {8\pi G}$ represents the
coupling constant term and $S_m$ is the matter Lagrangian.
The Gauss-Bonnet invariant term $\mathcal{G}$ is defined as
\begin{equation}\label{6}
\mathcal{G} = R^2 - 4R_{\mu\nu}R^{\mu\nu} +R_{\mu\nu\sigma\rho}R^{\mu\nu\sigma\rho},
\end{equation}
where $R_{\mu\nu}$ and $R_{\mu\nu\rho\sigma}$ indicate the Ricci and Riemann tensors, respectively.
Varying the action (\ref{5}) with respect to metric tensor $g_{\mu\nu}$, the
modified field equations turn out to be
\begin{eqnarray}\nonumber
G_{\mu\nu}+ 8\big[R_{\mu\rho\nu\sigma} + R_{\rho\nu}g_{\sigma\mu} - R_{\rho\sigma}g_{\nu\mu} - R_{\mu\nu}g_{\sigma\rho} + R_{\mu\sigma}g_{\nu\rho} + \frac{R}{2}(g_{\mu\nu}g_{\sigma\rho}-g_{\mu\sigma}g_{\nu\rho})\big]\nabla^{\rho}\nabla^{\sigma}f_\mathcal{G}+(\mathcal{G}f_\mathcal{G}- f)g_{\mu\nu} =\kappa^2T_{\mu\nu},\\\label{fe}
\end{eqnarray}
where the subscript $\mathcal{G}$ in $f_\mathcal{G}$ represents the derivative with respect to $\mathcal{G}$ and $T_{\mu\nu}$ is the stress-energy tensor defined as
\begin{equation}\label{8}
T_{\mu\nu}=(\rho+p_t)u_\mu u_\nu-p_tg_{\mu\nu}+(p_r-p_t)v_\mu v_\nu,
\end{equation}
where $u_\mu=e^{\nu/2} \delta_\alpha^0$, $v_\nu=e^{\lambda/2}\delta_\alpha^1$ are four velocity vectors.
We can re-write modified field equations (\ref{fe}) in an alternative form familiar with GR as
\begin{eqnarray}\label{teff}
G_{\mu\nu} =\kappa^2T_{\mu\nu}^{eff},
\end{eqnarray}
where effective stress-energy tensor $T_{\mu\nu}^{eff}$ is given by
\begin{eqnarray}\label{70}
T_{\mu\nu}^{eff}=T_{\mu\nu}- \frac{8}{\kappa^2}\big[R_{\mu\rho\nu\sigma} + R_{\rho\nu}g_{\sigma\mu} - R_{\rho\sigma}g_{\nu\mu} - R_{\mu\nu}g_{\sigma\rho} + R_{\mu\sigma}g_{\nu\rho} + \frac{R}{2}(g_{\mu\nu}g_{\sigma\rho}-g_{\mu\sigma}g_{\nu\rho})\big]\nabla^{\rho}\nabla^{\sigma}f_\mathcal{G}
 -(\mathcal{G}f_\mathcal{G}- f)g_{\mu\nu}.
\end{eqnarray}
It is interesting to notice that effective energy-momentum tensor consist of usual matter contents and also the matter contents from
geometric origin. Thus, it seems interesting as this approach may provide all the matter components which may be essential to unveil the phenomenon of dark energy and accelerated expansion.
Here we consider the curvature tensor convention by taking the signature of the Riemannian metric as $(+,-,-,-)$ and furthermore Riemann tensor
and covariant derivative for a vector field is denoted by $R^{\sigma}_{\mu\nu\rho}=\partial_{\nu}\Gamma^{\sigma}_{\mu\rho}-\partial_{\rho}\Gamma^{\sigma}_{\mu\nu}+
\Gamma^{\omega}_{\mu\rho}\Gamma^{\sigma}_{\omega\nu}-\Gamma^{\omega}_{\mu\nu}\Gamma^{\sigma}_{\omega\rho}$ and $\nabla_{\mu}V_{\nu}=\partial_{\mu}V_{\nu}-\Gamma^{\lambda}_{\mu }V_{\lambda}$ respectively.
Further, to examine and investigate the configurations of compact stars, we will choose the spacetime which is static, non-rotating and spherically symmetric every where \cite{Krori}.
\begin{equation}\label{9}
ds^{2}= e^{\nu(r)}dt^2-e^{\lambda(r)}dr^2-r^2(d\theta^2+\sin^2 \theta d\phi^2).
\end{equation}
Using equations (\ref{70}) and (\ref{9}) and after some manipulations we acquire the following set of modified field equations for the anisotropic stellar system as
\begin{eqnarray}\label{10}
\rho^{eff}&&=~~\rho-8e^{-2\lambda}(f_\mathcal{GGG}\mathcal{G}'^{2}+f_\mathcal{GG}\mathcal{G}'')(\frac{e^{\lambda}-1}{r^2})
+4e^{-2\lambda}\lambda' \mathcal{G}'f_\mathcal{GG}(\frac{e^{\lambda}-3}{r^2})-(\mathcal{G}f_\mathcal{G}-f),
\end{eqnarray}
\begin{eqnarray}\label{11}
p_{r}^{eff}&&=~~ ~p_{r}-4e^{-2\lambda}\nu'\mathcal{G}'f_\mathcal{GG}(\frac{e^{\lambda}-3}{r^2})+
(\mathcal{G}f_\mathcal{G}-f),\quad\quad\quad\quad\quad\\\nonumber
\end{eqnarray}
\begin{eqnarray}\label{12}
p_{t}^{eff}&&=~~p_{t}-\frac{4e^{-2\lambda}\nu'}{r}(f_\mathcal{GGG}\mathcal{G}'^{2}+f_\mathcal{GG}\mathcal{G}'')-\frac{2e^{-2\lambda}{\nu'}^{2}f_\mathcal{GG}\mathcal{G}'}{r}-\frac{2e^{-2\lambda}f_\mathcal{GG}\mathcal{G}'}{r}(2\nu''-3\nu'\lambda')+
(\mathcal{G}f_\mathcal{G}-f).
\end{eqnarray}
Here $\rho$, $p_r$ and $p_t$ are usual energy density, radial pressure and transverse pressure respectively.
The system of three equations have five unknown functions namely, $\rho^{eff}$, $p^{eff}_{r}$, $p^{eff}_{t}$, $\lambda$, $\nu$.
For the above equations of motion, the expressions for $\rho^{eff}$, $p^{eff}_{r}$ and $p^{eff}_{t}$ are equal to the components of the Einstein tensor.
\\\\$\mathbf{\textit{\textbf{Theorem}:}}$
Given a solution of Eqs. (\ref{10})-(\ref{12}), defined by the functions $T_1=\big\{\nu(r), \lambda(r), f(\mathcal{G})\big\}$, if we have a solution in GR defined by $T_2=\big\{\nu(r), \lambda(r)\big\}$, then all the physical attributes are identical for $T_1$
and $T_2$ since $T^{eff}_{\mu\nu}$ in (\ref{teff}) plays the role of stress-energy tensor in GR \cite{M.V}.
\\\\The Eqs. (\ref{10})-(\ref{12}) are very much intricate and non-linear because of the involved variable function $f(\mathcal{G})$. Here we consider  $\nu=Br^2+2lnC$ and $\lambda=ln(1 +
ar^2+br^4)$ with constant parameters $a$, $b$, $C$ and $B$. In this way the above spacetime (\ref{9}) with specified metric potentials is known as Tolman-Kuchowicz  spacetime \cite{Jasim}. In order to examine the structure and stability of compact stars, we consider viable $f(\mathcal{G})$ gravity models which enable us to compute the effective energy density $\rho^{eff}$, radial pressure $p^{eff}_{r}$ and  transverse pressure $p^{eff}_{t}$.
\section{The Realistic Viable $f(\mathcal{G})$  Gravity Models}
In this section we give the analysis of compact stars by using the two realistic  $f(\mathcal{G})$ gravity models.
\subsection{Model $\mathbf{1}$}
First, we consider a power-law model with the additional logarithmic correction term \cite{ Schmidt}
\begin{eqnarray}\label{13}
f_{1}=\alpha_1\mathcal{G}^{n_1}+\beta_1\mathcal{G}ln(\mathcal{G}),
\end{eqnarray}
where $\alpha_1$, $\beta_1$ and $n_1$ are arbitrary constants to be estimated depending on several physical requirements. Observationally well-consistent cosmic results are obtained by this model due to its extra degrees of freedom allowed in the dynamics \cite{Setare}.
To probe the possible existence of the compact stars for the model (\ref{13}), the values of constants are picked in such a way that effective energy density, effective pressure and all energy conditions remain positive for the given model under investigation. By making use of Eq.(\ref{13}), the explicit relation for the effective energy density, effective radial pressure and transverse pressure have been found to be:

\begin{eqnarray}\nonumber
&&\rho^{eff}=\rho+\frac{8}{(1 +ar^2 + br^4)^3 \mathcal{G}^{3}}\bigg[(a+2br^2)(ar^2+br^4-2)\mathcal{G}(\beta_1\mathcal{G} +( n_1-1 )n_1 \alpha_1 \mathcal{G}^{n_1})\mathcal{G}'-(a\\\nonumber
&&+br^2)(1+ar^2+br^4)(\beta_1\mathcal{G} (-\mathcal{G}'^2+ \mathcal{GG''})+( n_1-1) n_1 \alpha_1\mathcal{G}^{n_1}(( n_1-2 )\mathcal{G}'^2+  \mathcal{GG''}))\bigg]\\\label{14}
&&+\alpha_1\mathcal{G}^{n_1}-n_1\alpha_1\mathcal{G}^{n_1}-\beta_1\mathcal{G}(1+ln(\mathcal{G}))
 +\beta_1 \mathcal{G}ln(\mathcal{G}),
\end{eqnarray}
\begin{eqnarray}\nonumber
&&p^{eff}_{r}=p_{r}+\frac{8B(a r^2 +br^4-2)}{r(1+ar^2+br^4)^2\mathcal{G}^2}\bigg[(\beta_1\mathcal{G}+(n_1-1)n_1\alpha_1\mathcal{G}^{n_1})\mathcal{G}'\bigg]-\alpha_1\mathcal{G}^{n_1}
-\beta_1\mathcal{G} ln(\mathcal{G})+ n_1\alpha_1\mathcal{G}^{n_1}\\\nonumber
&&-\beta_1\mathcal{G}(1+ln(\mathcal{G})),~~~~~~~~~~~~~~~\\\label{15}
\end{eqnarray}

\begin{eqnarray}\nonumber
&&p^{eff}_{t}=p_{t}-\frac{1}{r (1 + a r^2 + b r^4)^3}r\mathcal{G}^{3}\bigg[-8B^2r^2(1 + a r^2 + b r^4)\mathcal{G}(\beta_1\mathcal{G}+( n_1-1)\alpha_1\mathcal{G}^{n_1})\mathcal{G}' \\\nonumber
&&+ 8B (2a r^2 + 5b r^4-1)\mathcal{G}(\beta_1\mathcal{G}+(n_1-1)n_1\alpha_1\mathcal{G}^{n_1}\mathcal{G})\mathcal{G'}+r(1+ar^2 + br^4)(8 B\beta_1\mathcal{G}(\mathcal{G}'^2+  \mathcal{GG''})\\\nonumber
&&-8B(n_1-1)n_1\alpha_1\mathcal{G}^{n_1}((n_1-2)\mathcal{G}'^2+ \mathcal{GG''}))\bigg]-\alpha_1\mathcal{G}^{n_1}+ n_1\alpha_1\mathcal{G}^{n_1}-\beta_1\mathcal{G}ln(\mathcal{G})+\beta_1\mathcal{G}(1+ln(\mathcal{G})). \\\label{16}
\end{eqnarray}

\subsection{Model $2$}
Next, we consider the realistic $f(\mathcal{G})$ gravity model \cite{Bamba}, which reproduce the current cosmic acceleration, namely
\begin{eqnarray}\label{17}
f_{2}= \alpha_2\mathcal{G}^{n_2}(\beta_2 \mathcal{G}^{m}+1),
\end{eqnarray}
where $\alpha_2$, $\beta_2$, $m$ and $n_2$ are arbitrary constants, and $n_2>0$. The model regard to $f_{2}$ is considered worthwhile
for the treatment of the finite time future singularities \cite{Noj3}. The physical features of the compact stars for the model (\ref{17}) by  using Eqs. (\ref{10})-(\ref{12}) can be defined by the following relation as

\begin{eqnarray}\nonumber
&&\rho^{eff}=\rho + \frac{8\mathcal{G}^{(n_2-3)}}{ r(1 + a r^2 + b r^4)^3}\bigg[(a+
2br^2)( a r^2 + b r^4-2 )\alpha_2\mathcal{G} (( n_2-1 ) n_2 + ( m + n_2-1 ) (m + n_2)\beta_2\mathcal{G}^m)
  \mathcal{G}' \\\nonumber
&&  -r^3(a+br^2)(1 + a r^2 + b r^4)(((n_2-2)(n_2-1)n_2\alpha_2+((n_2-2)(n_2-1)n_2\alpha_2+(m-2)(m-1)m\alpha_2\\\nonumber
&&+3mn_2(m+n_2-2)\alpha_2)\beta_2\mathcal{G}^m)\mathcal{G}'^2+
\alpha_2 \mathcal{G}((n_2-1)n_2+(m+n_2-1)(m+n_2)\beta_2 \mathcal{G}^m) \mathcal{G}''\bigg]+\alpha_2 \mathcal{G}^{n_2} (1 + \beta_2 \mathcal{G}^{m})\\\nonumber
&& - \alpha_2\mathcal{G}^
  {n_2} (n_2 + (m + n_2)\beta_2 \mathcal{G}^m),\\\label{18}
\end{eqnarray}
\begin{eqnarray}\nonumber
&&p^{eff}_{r}=p_{r}+\frac{8B(ar^2+br^4-2)\alpha_2\mathcal{G}^{(n_2-2)}}{r (1 + a r^2 + b r^4)^2}\bigg[(n_2-1)n_2 + (m^2+(n_2-1)n_2+m(2n_2-1))\beta_2 \mathcal{G}^m\bigg]\mathcal{G}'-\alpha_2 \mathcal{G}^{n_2} (1\\\label{l9}
 && + \beta_2 \mathcal{G}^{m}) + \alpha_2\mathcal{G}^
  {n_2} (n_2 + (m + n_2)\beta_2 \mathcal{G}^m),\\\nonumber
\end{eqnarray}
\begin{eqnarray}\nonumber
&&p^{eff}_{t}=p_{t}+\frac{8B\mathcal{G}^{(n_2-3)}}{ r(1 + a r^2 + b r^4)^3}\bigg[Br^2(1 + a r^2 +
   b r^4)\alpha_2\mathcal{G}((n_2-1)n_2+(m+n_2-1)(m +n_2)\beta_2 \mathcal{G}^m)\mathcal{G}' \\\nonumber
   &&+(2ar^2+5br^4-1)\alpha_2\mathcal{G}((n_2-1)n_2+(m+n_2-1)(m+n_2)\beta_2 \mathcal{G}^m)\mathcal{G}')-r(1 + a r^2 + b r^4)(((n_2\\\nonumber
  &&-2)(n_2-1)n_2\alpha_2+((n_2
 -2)(n_2-1)n_2\alpha_2+(m-2)(m-1)m\alpha_2
+3mn_2(m+n_2-2)\alpha_2)\beta_2\mathcal{G}^m)\mathcal{G}'^2\\\label{20}
&&+\alpha_2 \mathcal{G}((n_2-1)n_2+(m+n_2-1)(m+n_2)\beta_2 \mathcal{G}^m) \mathcal{G}''\bigg]
-\alpha_2 \mathcal{G}^{n_2} (1 + \beta_2 \mathcal{G}^{m})+ \alpha_2\mathcal{G}^{n_2} (n_2 + (m + n_2)\beta_2 \mathcal{G}^m).\\\nonumber
   \end{eqnarray}

\subsection{Boundary Conditions}
\begin{figure}[h!]
\begin{tabular}{cccc}
\epsfig{file=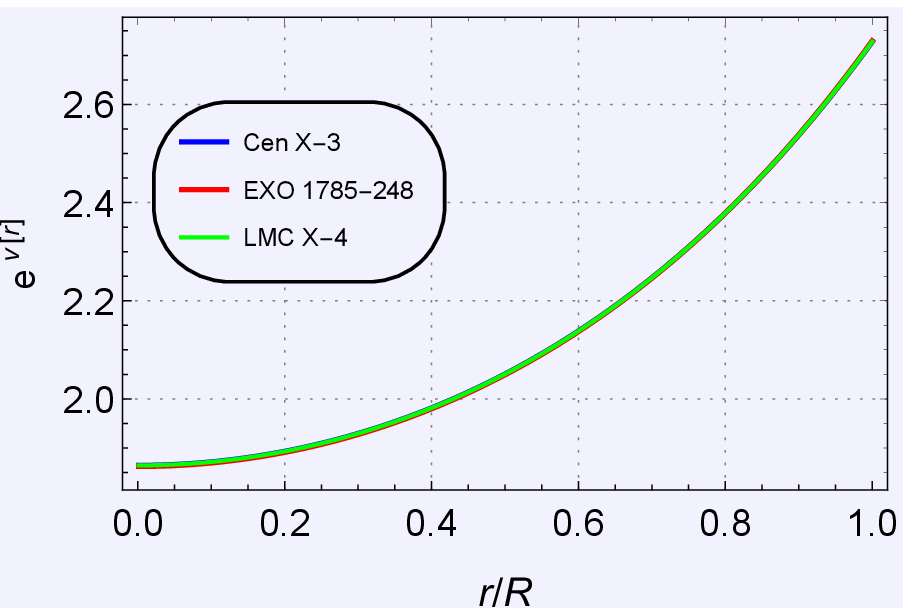,width=0.38\linewidth} &
\epsfig{file=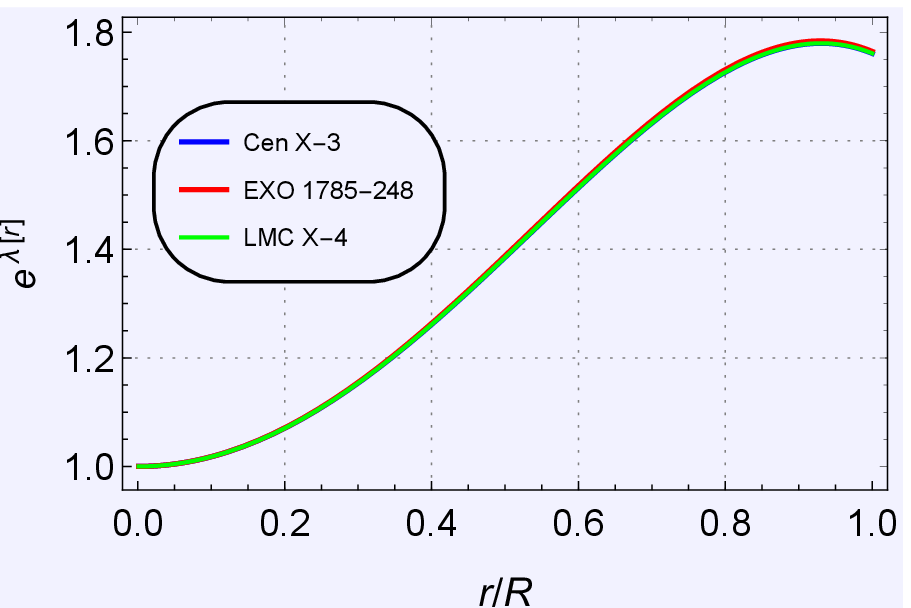,width=0.38\linewidth} &
\end{tabular}
\caption{Behavior of metric potentials for stars, $Cen X-3$, $EXO 1785-248$ and $LMC X-4$.}
\label{Fig:1}
\end{figure}

The existence of physical and geometric
singularities within the star is considered as the most important features in the study of compact stellar objects.
For the existence of singularises,  we analyze the behavior of both metric potentials  $e^{\nu(r)}$ and $e^{\lambda(r)}$
at the center of structure $r=0$. For a physical viability and stability of the model, the metric potentials should be singularity-free, positive, monotonically increasing and regular inside the compact stellar structure.
The variation of metric potentials at the center of the star i.e. $e^{\lambda(r=0)}=1$  and $e^{\nu(r=0)}=C^2$ is shown in Fig. $\ref{Fig:1}$.
It is observed that the considered election on metric potentials are consistent with the above-mentioned conditions.
The graphical behavior shows that value of the both metric potentials at the center is minimum, then it increase nonlinearly and become maximum at the boundary surface.
\FloatBarrier
Further, it is important to mention here for the well behaved compact stellar objects, the following conditions should be satisfied
\begin{itemize}
\item $c^2\rho^{eff}$ should always be greater than $p^{eff}$ within the range $0\leq r\leq R.$
\item At the surface boundary $r=R$, the radial pressure must be zero i.e $p^{eff}_r(r=R)=0$.
    \item The effective density gradient $\frac{d^{eff}\rho}{dr}$ must be negative for the range $0\leq r\leq R$, i.e. $(\frac{d\rho^{eff}}{dr})_{r=0}$ and $(\frac{d^2p^{eff}}{dr^2})_{r=0}<0$.

 \item The effective pressure gradient $\frac{dp^{eff}_{r}}{dr}$ must be negative for the range $0\leq r\leq R$, i.e. $(\frac{dp^{eff}_{r}}{dr})_{r=0}$ and $(\frac{d^2p^{eff}_{r}}{dr^2})_{r=0}<0$.
These above two conditions depict that the effective energy density and effective radial pressure should be decreasing towards the boundary of the surface of the structure.

\item The velocity of sound speed must not be exceed the speed of light i.e. $\frac{dp^{eff}}{c^2d\rho}<1$.

\item The adiabatic index $\Gamma=\frac{\rho^{eff}+p^{eff}_{r}}{p^{eff}_r}(\frac{dp^{eff}_{r}}{d\rho^{eff}})= \frac{\rho^{eff}+p^{eff}_{r}}{p^{eff}_r}v^2_{r}>4/3$, is necessary condition for the stability.
\item The surface redshift $z_{s}$ must be finite and positive.

\end{itemize}
Here, we assume $c=1$. These physical attributes like  effective  energy density, pressure, sound speed  mass, surface redshift are the most significant features describing the structure of compact star. Now to check the appropriated behaviour and capability of characterizing the realistic stars, we plot the graphs of these features.
\section{Exterior Metric and Matching Conditions}

The intrinsic boundary metric remains the same, whether it is constructed from the interior and exterior geometry of the star. This mechanism ensures that the metric components irrespective of the coordinate system across
the boundary surface will remain continuous. No doubt, in theory of GR the Schwarzschild's solution is considered as the appropriate choice to choose from the diverse possibilities of the matching conditions while exploring the compact stellar objects. Also according to the Jebsen-Birkhoff's theorem statement, every spherically symmetric vacuum solution of field equations must be static and asymptotically flat.
Furthermore, as concern with the modified $f(\mathcal{G})$ gravity, the Schwarzschild's solution may be accommodated with a proper choice of viable
$f(\mathcal{G})$ gravity models for non zero density and pressure. Perhaps, this fact leads to the violation of Birkhof's theorem in modified theories of gravity \cite{Faraoni}. A lot of work on matching conditions has been done by many authors (\cite{Abbas2}-\cite{Abbas3},\cite{Bhar}). The junction conditions that appear in the extended theories of gravity implement some restrictions on the stellar
objects are proved by Goswami et al. \cite{Goswami}. However for this goal many authors \cite{Cooney}-\cite{Ast1} have considered the Schwarzschild's solution, giving some fascinating results. At this juncture to solve the field equations under the specified constraint at $r=R$, the pressure $p_r(r=R)=0$, we match the intrinsic metric (\ref{9}) to the vacuum Schwarzschild's exterior metric, given by
\begin{equation}\label{18a}
ds^2=(1-\frac{2M}{r}) dt^2 -(1-\frac{2M}{r})^{-1} dr^2 -r^2 d\theta^2 -r^2\sin^2\theta d\phi^2,
\end{equation}
where ``$M$" stands for the total mass within the boundary of the compact star. At the boundary surface $r = R$, the continuity of the metric potentials yield the following expressions
\begin{equation}\label{19a}
g^{-}_{tt} = g^{+}_{tt}, ~~~~ g^{-}_{rr} = g^{+}_{rr},~~~~\frac{\partial g^{-}_{tt}}{\partial r} = \frac{\partial g^{+}_{tt}}{\partial r},
\end{equation}
where interior and exterior solutions are symbolized by $(-)$
and $(+)$ respectively. The values of the constants $a,~b,~B$ and $C$ are obtained by comparing the interior and exterior of the
metric such as
\begin{equation}\label{20a}
a=\frac{1}{R(R-2M)}+\frac{MR}{2(R-2M)^4}-\frac{1}{R^2},~~~~~~~~~~~b=\frac{-M}{2R(R-2M)^4},
\end{equation}
\begin{equation}\label{21a}
B=\frac{M}{R^3}(1-\frac{2M}{R})^{-1},~~~~~~~~~~~~~~~~~~~~~~ C=e^{\frac{1}{2}{[\ln(1-\frac{2M}{R})-\frac{M}{R}(1-\frac{2M}{R})^{-1}]}}.
\end{equation}
The approximated values of mass and radius of the compact stars $Cen~ X-3$, $EXO ~1785-248$ and $LMC~ X-4$ are
considered to find out these constant values $a,  b,~ B$ and $C$ which are given in the following Table $\ref{tab1}$.
\begin{table}[ht]
\caption{The approximated
values of unknown constants $a$, $b$, $B$ and $C$
for compact star candidates $Cen~ X-3$, $EXO~ 1785-248$ and $LMC~ X-4$.}
\centering
\begin{tabular}{|p{2.8cm}|p{3.3cm}| p{3.3cm}| p{3.3cm}|}
\hline
\hline
Star Model & ~~~ $Cen~ X-3$              &~~~ $EXO~ 1785-248$          &~~~ $LMC ~X-4$ \\
\hline
~~~~ $M$       &~~~ 1.49 $\pm$ 0.08 ~~\cite{Gango} &~~~  1.3 $\pm$ 0.2~~ \cite{Ozel}& ~~~  1.29 $\pm$ 0.05~~ \cite{Gango}    \\
\hline
~~~~ $R$       &~~~  9.508 $\pm$ 0.115             &~~~ 9.189 $\pm$ 0.396           &~~~   9.170 $\pm$ 0.098 \\
\hline
~~~~ $\mu=M/R$ &~~~ 0.2226                         & ~~~ 0.2166                      & ~~~  0.2160 \\
\hline
~~~~ $a$      &~~~  0.0224207                      &~~~  0.0197478                   & ~~~  0.0212739 \\
\hline
~~~~ $b$      &~~~  -0.000151008                   &~~~  -0.000124379                &~~~   -0.00014514 \\
\hline
~~~~ $B$      & ~~~  0.00454877                    &~~~ 0.0041604                    &~~~  0.00449728\\
\hline
~~~~ $C$      &~~~  0.609389                       & ~~~  0.621865                   &~~~  0.623022  \\
\hline
\end{tabular}
\label{tab1}
\end{table}
\FloatBarrier
\section{Physical Aspects of $f(\mathcal{G})$ Gravity Model}
In this section, we study various physical attributes of the anisotropic  dense stellar objects such as, effective energy density, radial and transversal pressure, energy bounds, anisotropic factor, mass function, compactification parameter and analysis of surface redshift and adiabatic index of our proposed models for the specific  values of the model parameters.
\subsection{Energy Density and Pressure Evolutions}
\begin{figure}[h!]
\begin{tabular}{cccc}
\epsfig{file=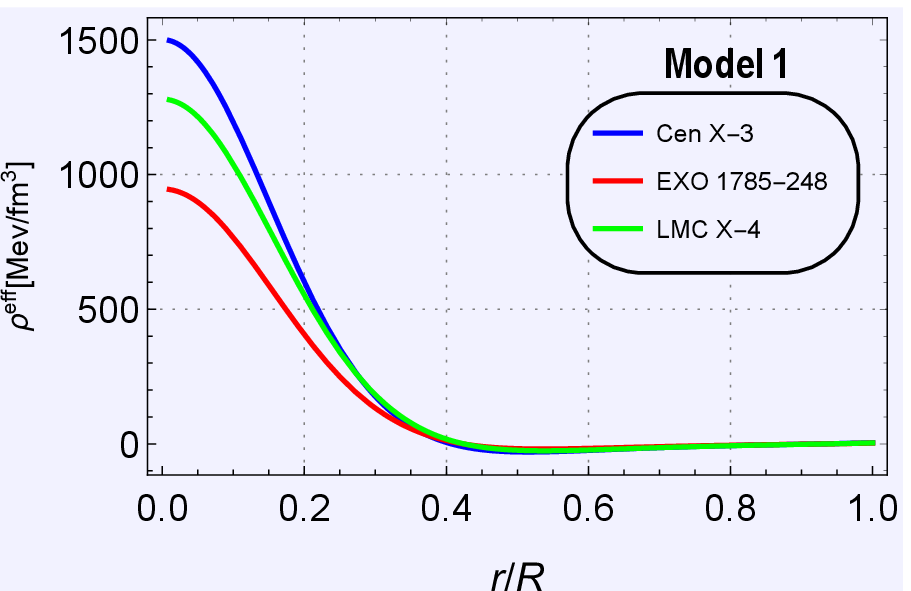,width=0.27\linewidth} &
\epsfig{file=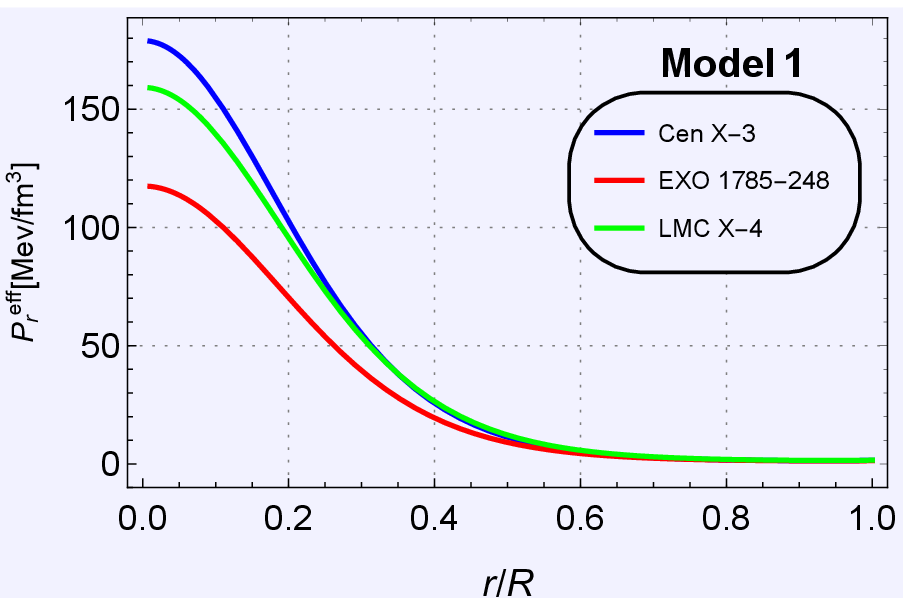,width=0.27\linewidth} &
\epsfig{file=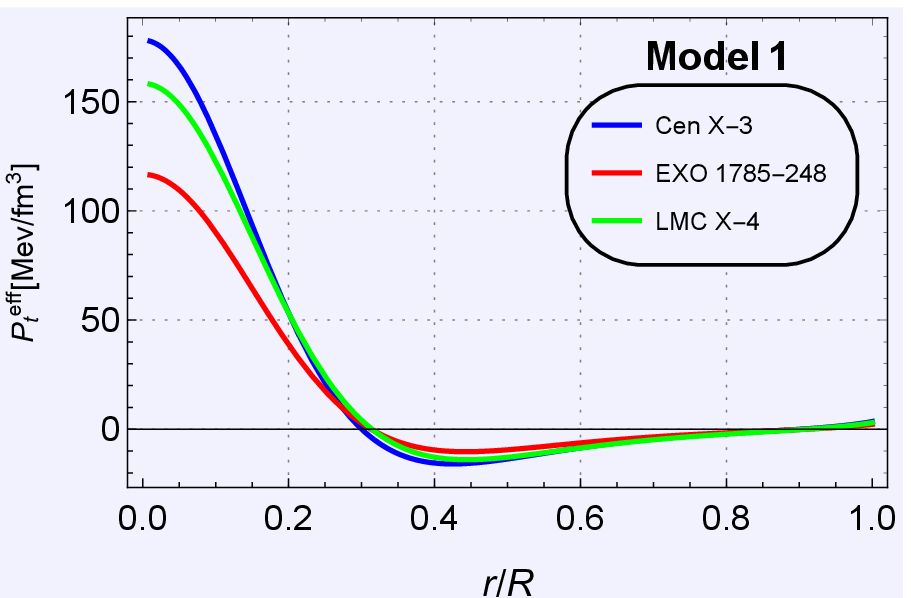,width=0.27\linewidth} &
\end{tabular}
\caption{Evolution of effective energy density(left panel), effective radial pressure (middle panel)and effective transverse pressure (right panel) for stars, $Cen X-3$, $EXO 1785-248$ and $LMC X-4$, under viable $f(\mathcal{G})$ gravity model $1$.}
\label{Fig:2}
\end{figure}

\begin{figure}[h!]
\begin{tabular}{cccc}
\epsfig{file=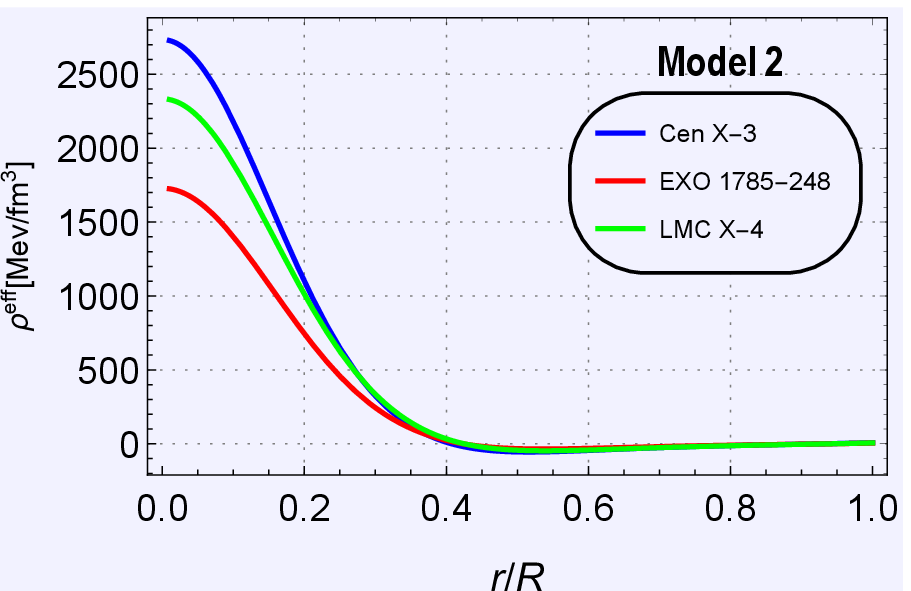,width=0.27\linewidth} &
\epsfig{file=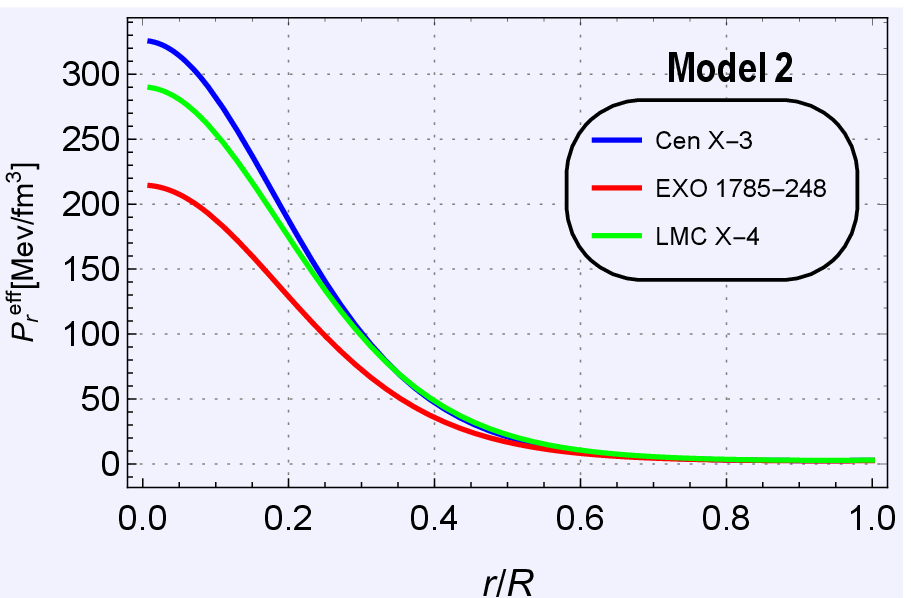,width=0.27\linewidth} &
\epsfig{file=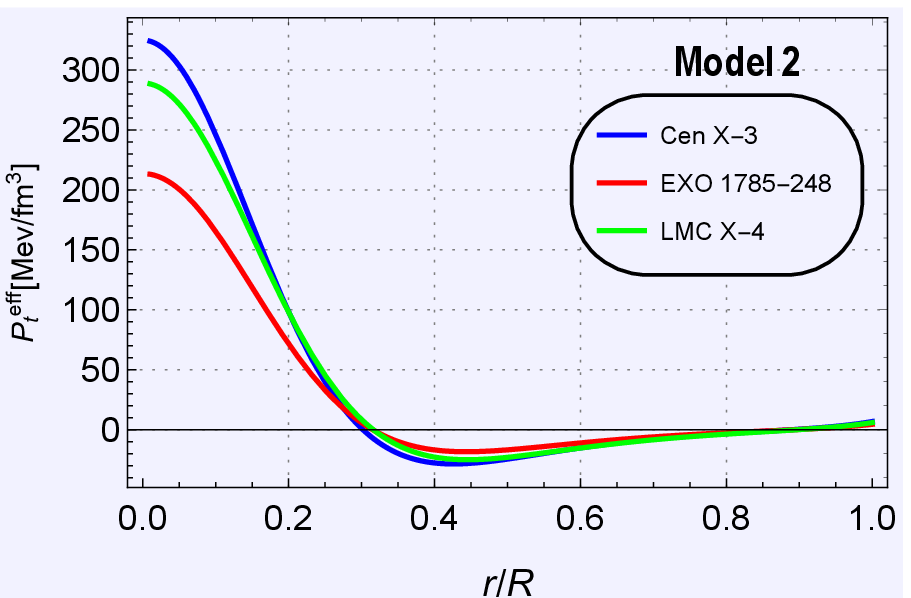,width=0.27\linewidth} &
\end{tabular}
\caption{Evolution of effective energy density(left panel), effective radial pressure (middle panel) and effective transverse pressure (right panel) for stars, $Cen X-3$, $EXO 1785-248$ and $LMC X-4$, under viable $f(\mathcal{G})$ gravity model $2$.}
\label{Fig:3}
\end{figure}

The effective energy density and pressure components
inside the stellar system show maximal value due to densest nature of compact relativistic objects.
The graphical analysis of the effective energy density, radial and transverse pressure
for considered compact star candidates with respect to
fractional radial coordinate $r/R$ is shown in Figs. $\ref{Fig:2}$ and $\ref{Fig:3}$.
\FloatBarrier
\FloatBarrier
From this graphical behavior it is clear that at the center of the compact stars, the effective energy density and components of pressure attain its maximum values and further approaches to the zero on the surface boundary, which point out this fact that our stellar objects show very high compactness.
These plots clearly demonstrate the existence of anisotropic configuration of compact stars for our suggested models in $f(\mathcal{G})$ gravity.
The numerical values of effective central density and radial pressure for three compact stars are shown in Table $\ref{tab2}$ and $\ref{tab3}$.
These physical features are positive finite at the center, which confirms that our present system is free from physical and geometrical singularities.


\begin{table}[H]
\caption{The numerical values of central density and pressure for the parameters $\rho=0.2$, $p_{r}=0.8$, $p_{t}=0.0011$, $n_1=2$, $\alpha_1=-3.64534*10^6$, $\beta_1=10$ under viable $f(\mathcal{G})$ gravity model $1.$}
\centering
\begin{tabular}{|p{2.7cm}|p{2.5cm}| p{2.5cm}| p{2.8cm}| p{2.8cm}| p{2.8cm}|}
\hline\hline
$Star Model$ &~~~ $M$              & ~~~ $R$            &~~~ $\rho_c ~(g/cm^3)$  &~~~ $p_r~~(dyne/cm^2)$ \\
\hline
$Cen~ X-3$ & ~~1.49 $\pm$ 0.08   &~~9.508 $\pm$ 0.115 &~~ $2.66509*10^{15}$ & ~~ $2.86296*10^{35}$  \\
\hline
$EXO~ 1785-248$ & ~~1.3 $\pm$ 0.2 & ~~9.189 $\pm$ 0.396 &~~ $1.68087*10^{15}$ &~~ $1.88015*10^{35}$  \\
\hline
$LMC ~X-4$ &~~1.29 $\pm$ 0.05     & ~~9.170 $\pm$ 0.098 &~~~$2.27270*10^{15}$ & ~~ $2.54677*10^{35}$  \\
\hline
\end{tabular}
\label{tab2}
\end{table}

\begin{figure}[h!]
\begin{tabular}{cccc}
\epsfig{file=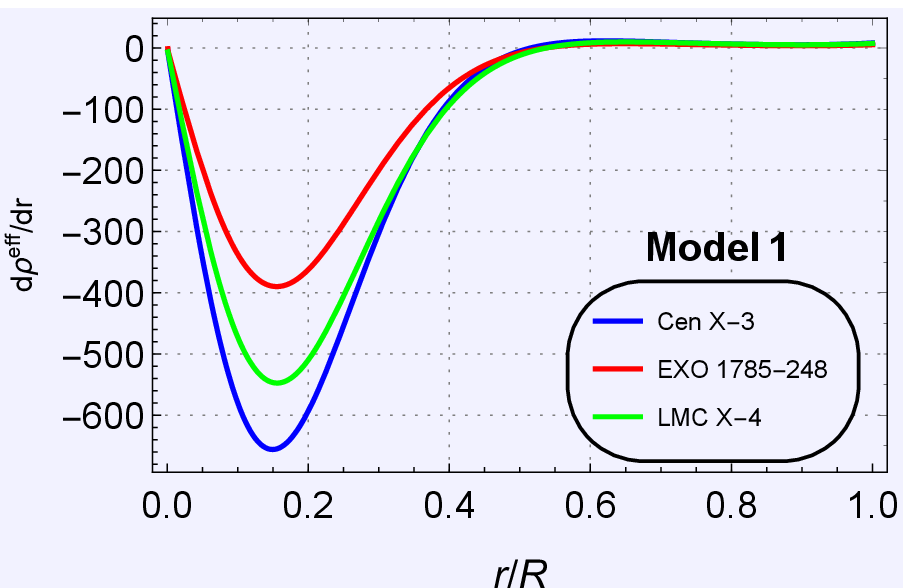,width=0.27\linewidth} &
\epsfig{file=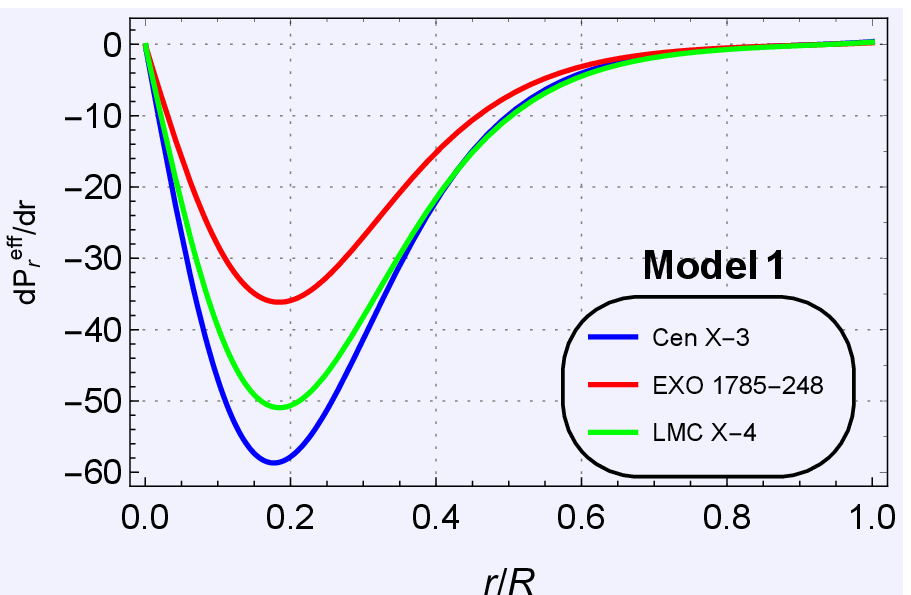,width=0.27\linewidth} &
\epsfig{file=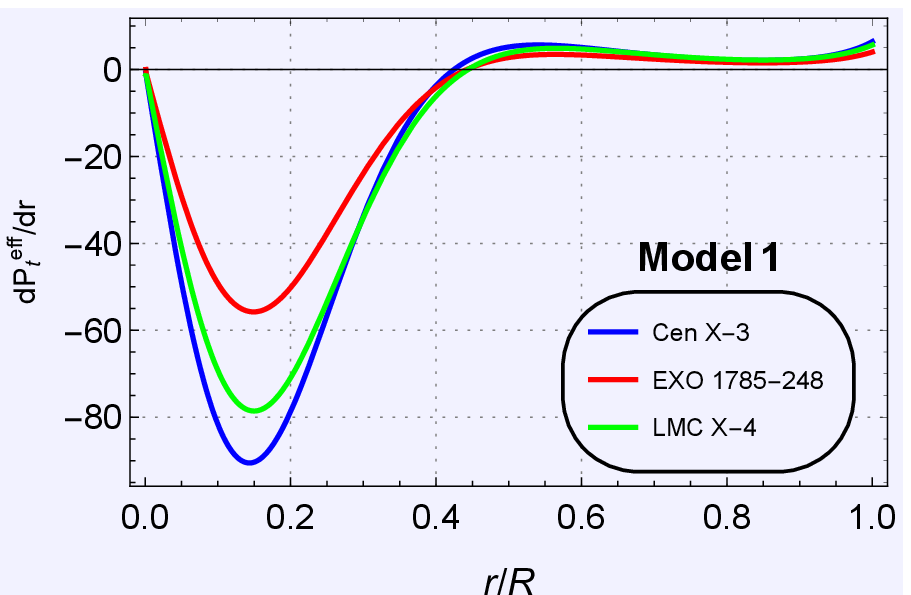,width=0.27\linewidth} &
\end{tabular}
\caption{Evolution of $\frac{d\rho^{eff}}{dr}$ (left panel), $\frac{dp^{eff}_r}{dr}$ (middle panel) and $\frac{dp^{eff}_t}{dr}$ (right panel) for stars, $Cen X-3$, $EXO 1785-248$ and $LMC X-4$, under viable  $f(\mathcal{G})$ gravity model $1$.}
\label{Fig4}
\end{figure}

\begin{figure}[h!]
\begin{tabular}{cccc}
\epsfig{file=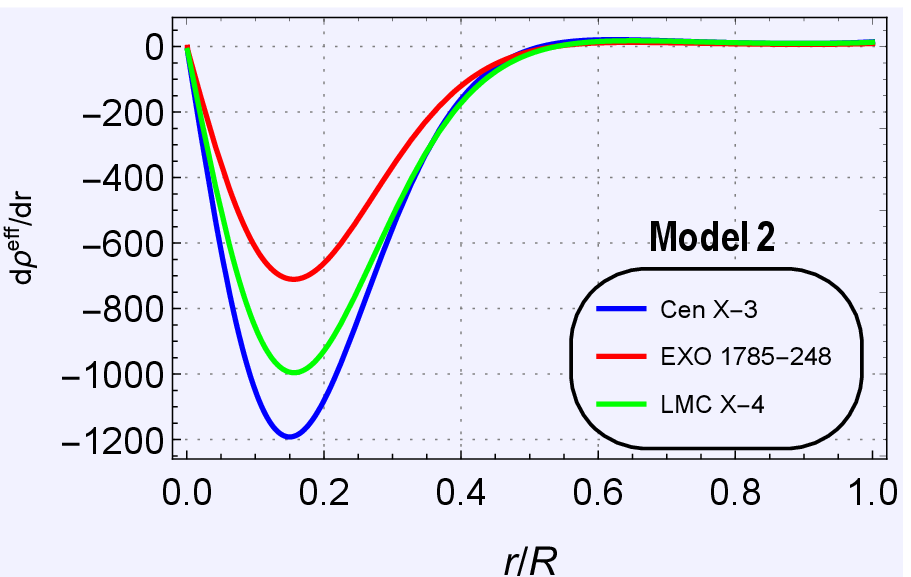,width=0.27\linewidth} &
\epsfig{file=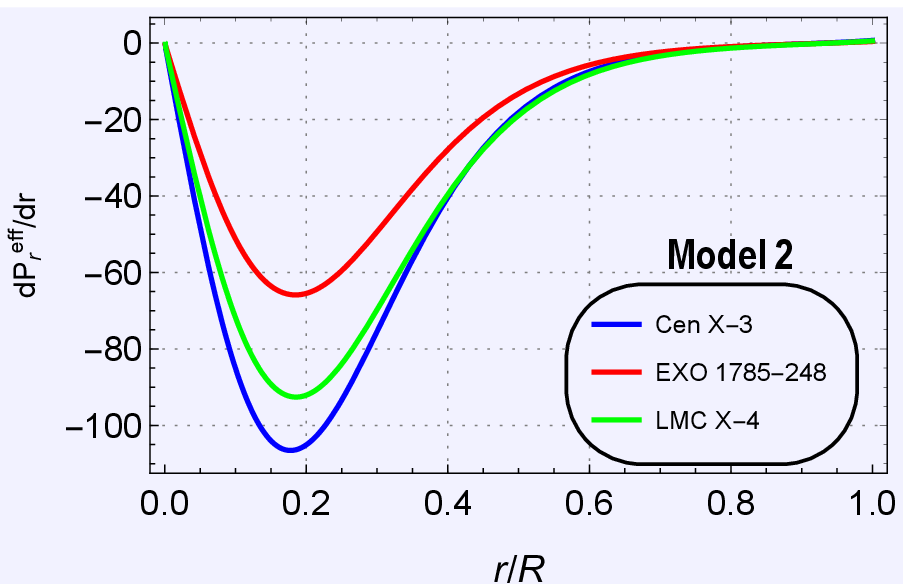,width=0.27\linewidth} &
\epsfig{file=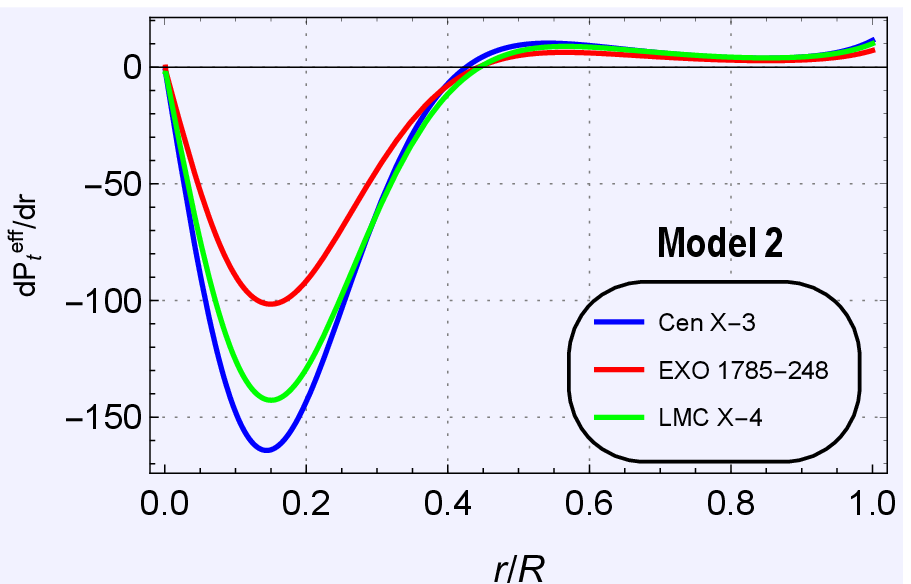,width=0.27\linewidth} &
\end{tabular}
\caption{Evolution of $\frac{d\rho^{eff}}{dr}$ (left panel), $\frac{dp^{eff}_r}{dr}$ (middle panel) and $\frac{dp^{eff}_t}{dr}$ (right panel) for stars, $Cen X-3$, $EXO 1785-248$ and $LMC X-4$, under viable $f(\mathcal{G})$ gravity model $2$.}
\label{Fig5}
\end{figure}

\begin{figure}[h!]
\begin{tabular}{cccc}
\epsfig{file=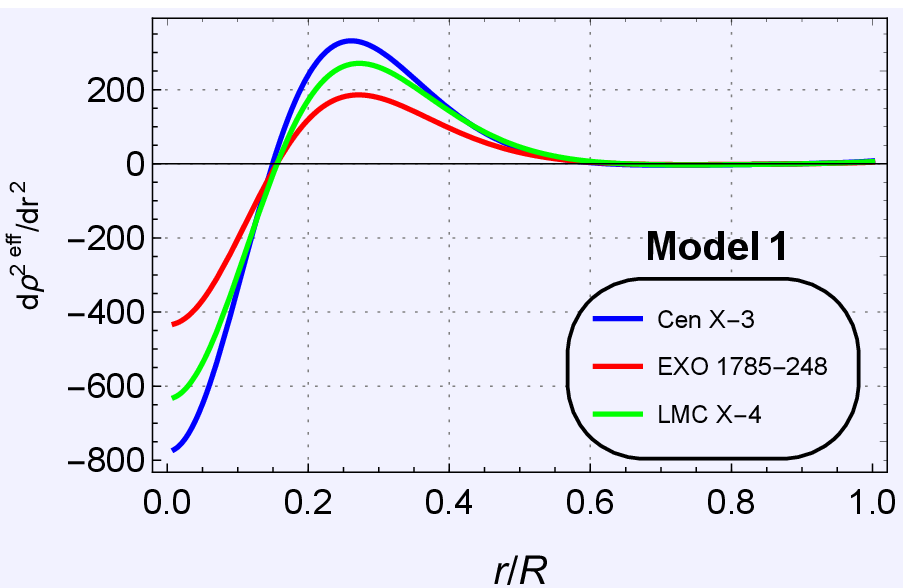,width=0.27\linewidth} &
\epsfig{file=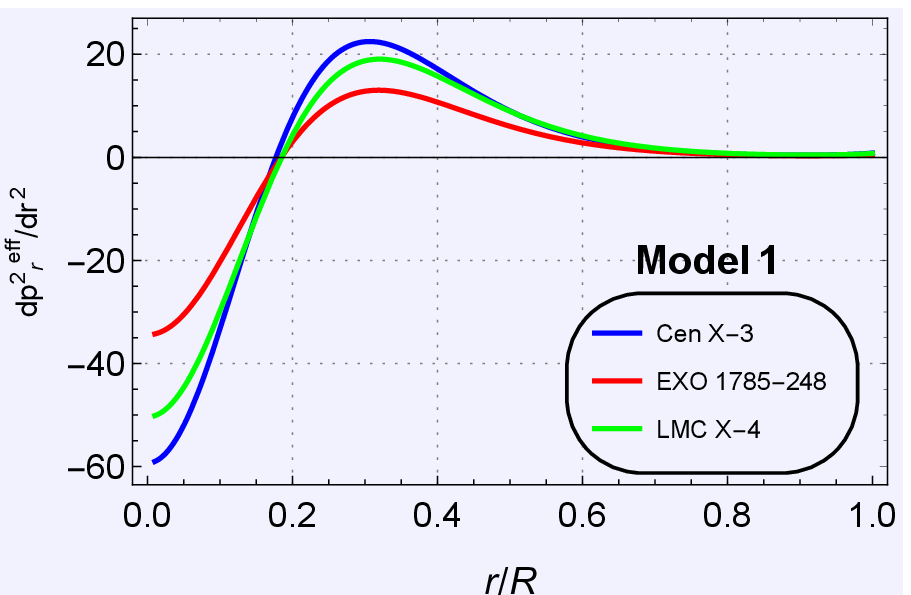,width=0.27\linewidth} &
\epsfig{file=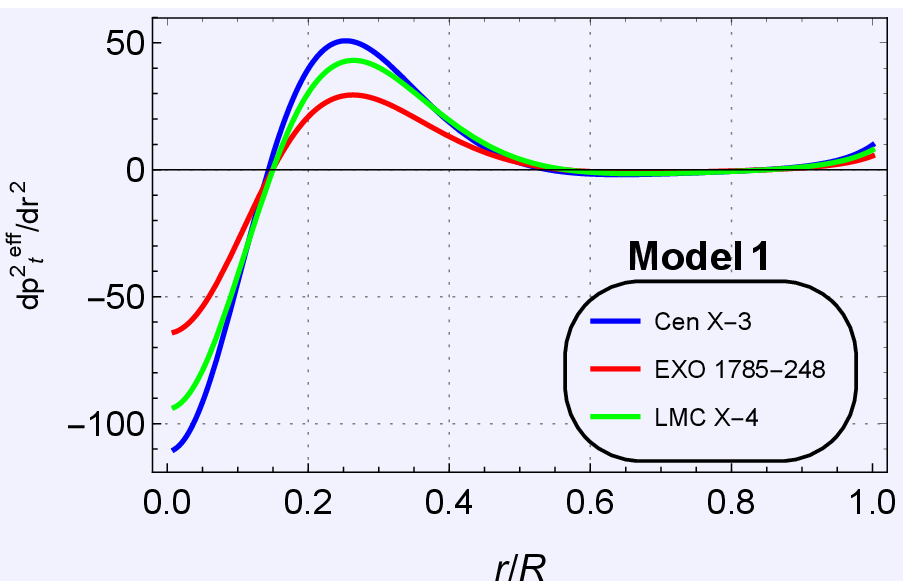,width=0.27\linewidth} &
\end{tabular}
\caption{Evolution of $\frac{d^2\rho^{eff}}{dr^2}$ (left panel), $\frac{d^2p^{eff}_r}{dr^2}$ (middle panel) and $\frac{d^2p^{eff}_t}{dr^2}$ (right panel) for  stars, $Cen X-3$, $EXO 1785-248$ and $LMC X-4$, under viable $f(\mathcal{G})$ gravity model $1$.}
\label{Fig6}
\end{figure}

\begin{figure}[h!]
\begin{tabular}{cccc}
\epsfig{file=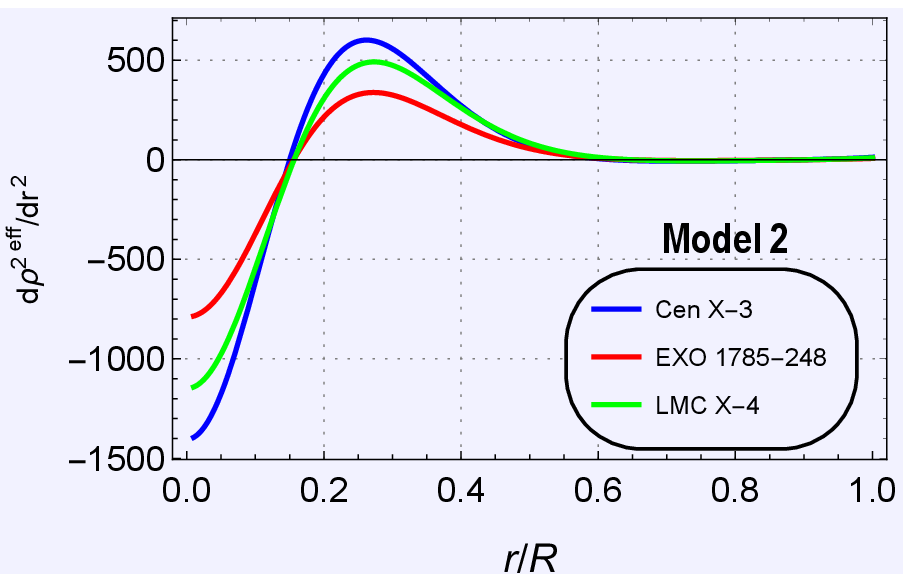,width=0.27\linewidth} &
\epsfig{file=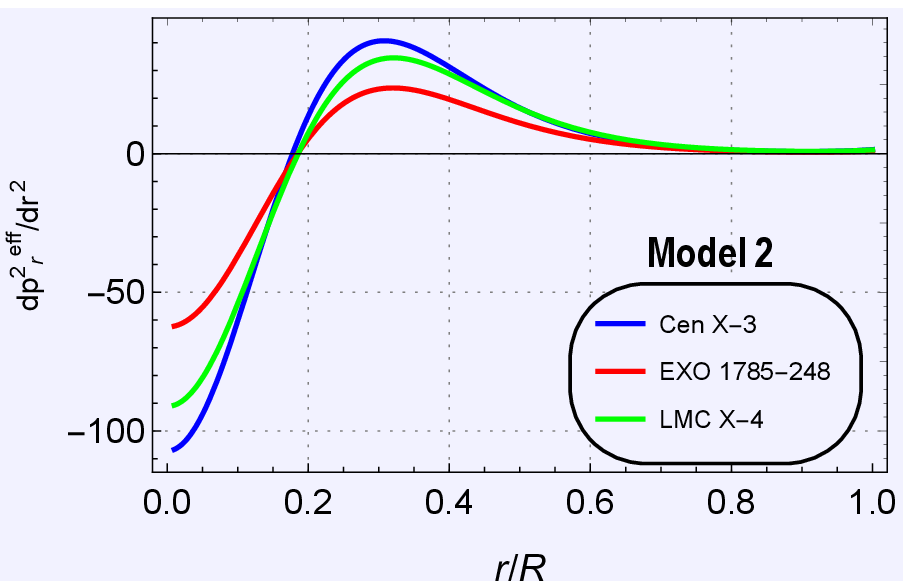,width=0.27\linewidth} &
\epsfig{file=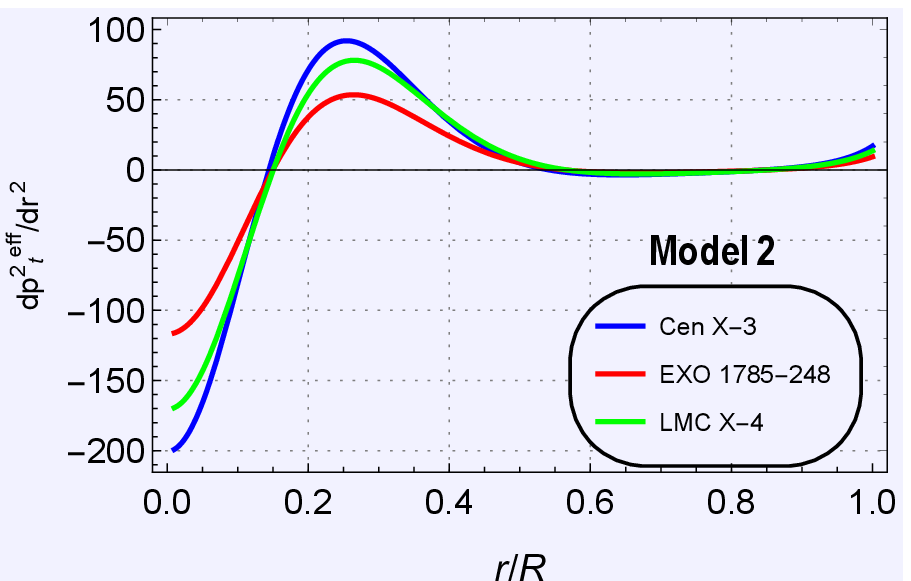,width=0.27\linewidth} &
\end{tabular}
\caption{Evolution of $\frac{d^2\rho^{eff}}{dr^2}$ (left panel), $\frac{d^2p^{eff}_r}{dr^2}$ (middle panel) and $\frac{d^2p^{eff}_t}{dr^2}$ (right panel) for stars, $Cen X-3$, $EXO 1785-248$ and $LMC X-4$, under viable $f(\mathcal{G})$ gravity model $2$.}
\label{Fig7}
\end{figure}

\begin{table}[H]
\caption{The numerical values of central density and  pressure for the parameters $\rho=0.2$, $p_{r}=1.5$, $p_{t}=0.5$, $n_2=2$, $m=1$, $\alpha_2=-6.74308*10^{-6}$, $\beta_2=2$ under viable  $f(\mathcal{G})$ gravity model $2$.}
\centering
\begin{tabular}{|p{2.7cm}|p{2.5cm}| p{2.5cm}| p{2.8cm}| p{2.8cm}| p{2.8cm}| }
\hline\hline
$Star Model$ & ~~~$M$              &~~~  $R$            & ~~~$\rho_c~(g/cm^3)$  & ~~~$p_r~~(dyne/cm^2)$ \\
\hline
$Cen~ X-3$      &~~ 1.49 $\pm$ 0.08    & ~~9.508 $\pm$ 0.115   &~~$4.85481*10^{15}$   & ~~~$5.2131*10^{35}$  \\
\hline
$EXO~ 1785-248$ & ~~~1.3 $\pm$ 0.2       &~~9.189 $\pm$ 0.396  &~~ $3.07022*10^{15}$  &~~ $3.43362*10^{35}$  \\
\hline
$LMC ~X-4$      &~~ 1.29 $\pm$ 0.05     & ~~9.170 $\pm$ 0.098  &~~ $4.14364*10^{15}$  &~~ $4.64173*10^{35}$  \\
\hline
\end{tabular}
\label{tab3}
\end{table}
The variation of the radial derivative of the effective energy density, radial and transverse pressure are denoted by  $\frac{d\rho^{eff}}{dr}$,
$\frac{dp^{eff}_{r}}{dr}$ and $\frac{dp^{eff}_{t}}{dr}$ respectively.
The graphical representation of these derivatives are shown in Figs. $\ref{Fig4}$ and $\ref{Fig5}$.
We observe that at the center $r=0$ these variations show decreasing evolution for the first order derivatives and expressed as
\begin{equation}\label{a}
\frac{d\rho^{eff}}{dr}<0,~~~~~~~~~~~~~~~~~~~~\frac{dp^{eff}_{r}}{dr}<0.
\end{equation}
\FloatBarrier
\FloatBarrier
It is also noticed that the second order derivatives of the density and radial  pressure at the center $r=0$ showing the maximum value, defined as
\begin{equation}\label{b}
\frac{d\rho^{eff}}{dr}=0=\frac{dp^{eff}_r}{dr},
~~~~~~~~~~\frac{d^2\rho^{eff}}{dr^2}<0,~~~~~~~~~~\frac{d^2p^{eff}_{r}}{dr^2}<0.
\end{equation}
The graphical analysis is shown in Figs. $\ref{Fig6}$ and $\ref{Fig7}$ which clearly depicts the compactness of the stars.
\FloatBarrier
\FloatBarrier
\subsection{Energy Conditions}
Some physical properties known as energy conditions are very helpful to investigate the presence of the realistic matter distribution.
Furthermore, these conditions play a vital role to identify the normal and exotic nature of matter inside the stellar structure model.
These energy conditions have captured much attention in
the discussion of some cosmological issues. By using the energy conditions one can easily examine
the validity of the second law of black hole thermodynamics and Hawking-Penrose singularity theorems \cite{Hawking}.
In cosmology many riveting results have been described by the use of energy conditions \cite{Santos}-\cite{Bertolami}.
These energy conditions are segregated into null, weak, strong and dominant
energy bounds and symbolized by NEC, WEC, SEC and DEC, respectively. These conditions in the presence of anisotropic fluid (\ref{9}) for curvature-matter coupled gravity \cite{Gasperini} are defined as
\begin{eqnarray}\nonumber
&&NEC:~~~~~~\rho^{eff}+p^{eff}_r\geq0,~~~\rho^{eff}+p^{eff}_t\geq0\nonumber,
\\&&WEC:~~~~~~\rho^{eff}\geq0,~~~\rho^{eff}+p^{eff}_r\geq0,~~~\rho+p^{eff}_t\geq0\nonumber,
\\&&SEC:~~~~~~\rho^{eff}+p^{eff}_r\geq0,~~~\rho^{eff}+p^{eff}_t\geq0,~~~\rho^{eff}+p^{eff}_r+2p^{eff}_t\geq0\nonumber,
\\&&DEC:~~~~~~\rho^{eff}-p^{eff}_r\geq0,~~~\rho^{eff}-p^{eff}_t\geq0.
\end{eqnarray}
All energy conditions have been satisfied for our chosen $f(\mathcal{G})$ models, as represented graphically
in Figs. $\ref{Fig:8}$ and $\ref{Fig:9}$
\begin{figure}[h!]
\begin{tabular}{cccc}
\epsfig{file=densitym1.eps,width=0.27\linewidth} &
\epsfig{file=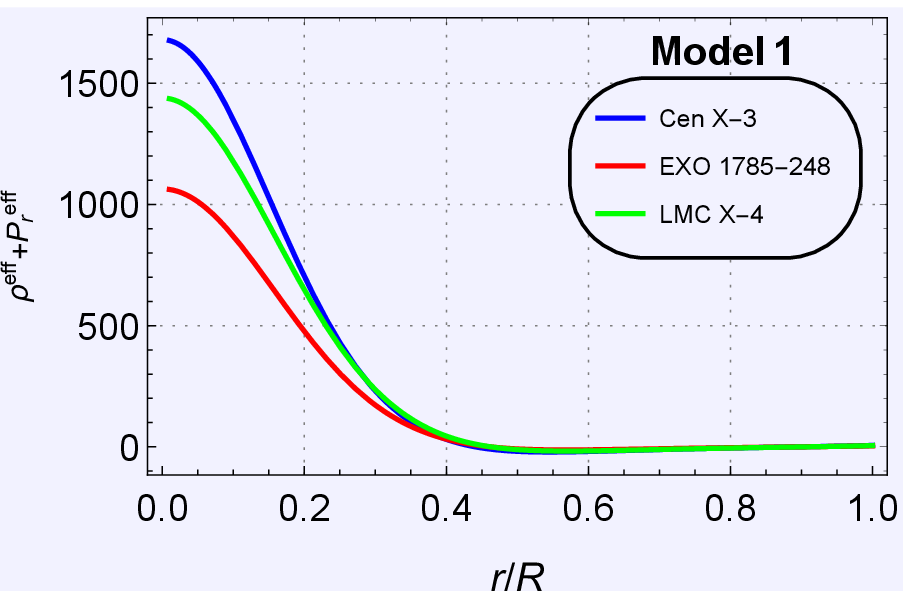,width=0.27\linewidth} &
\epsfig{file=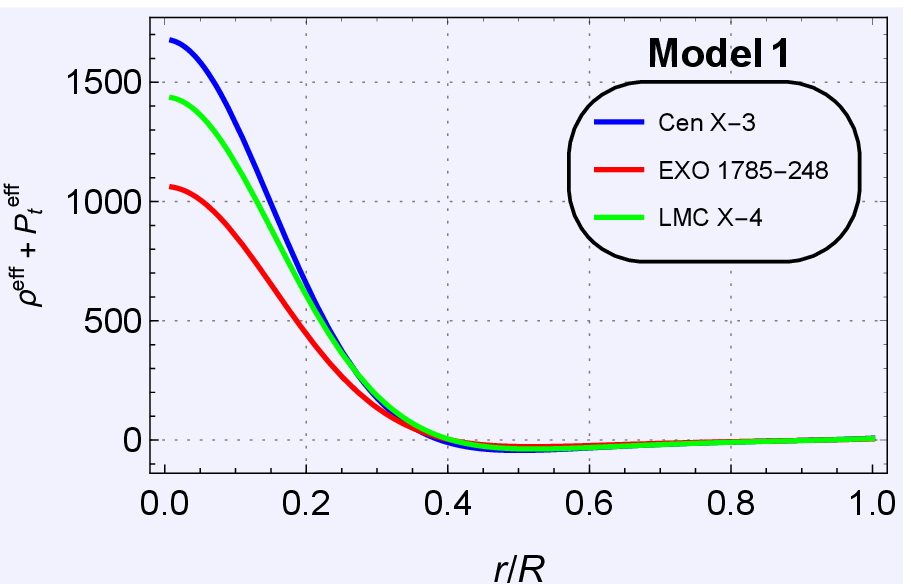,width=0.27\linewidth} &
\end{tabular}
\end{figure}
\FloatBarrier
\begin{figure}[h!]
\begin{tabular}{cccc}
\epsfig{file=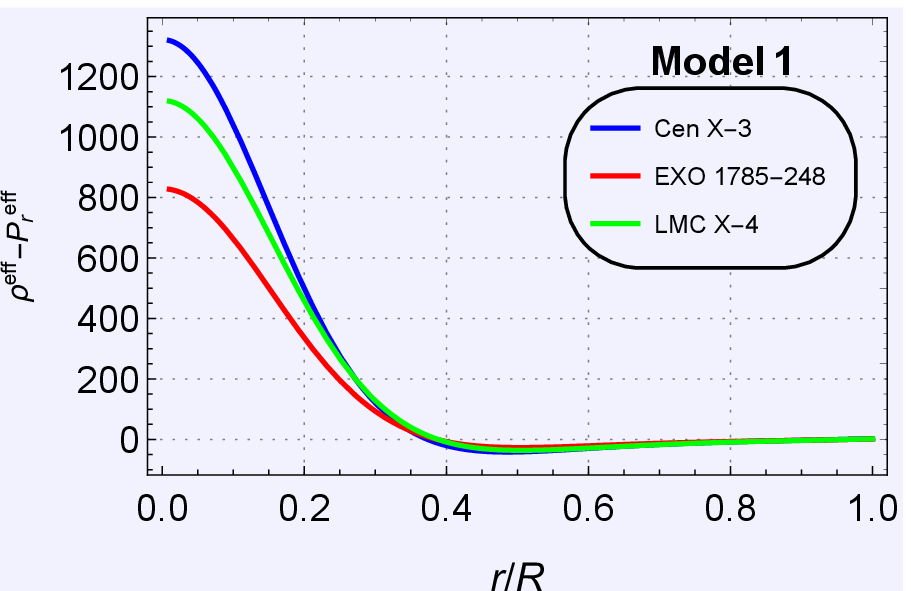,width=0.27\linewidth} &
\epsfig{file=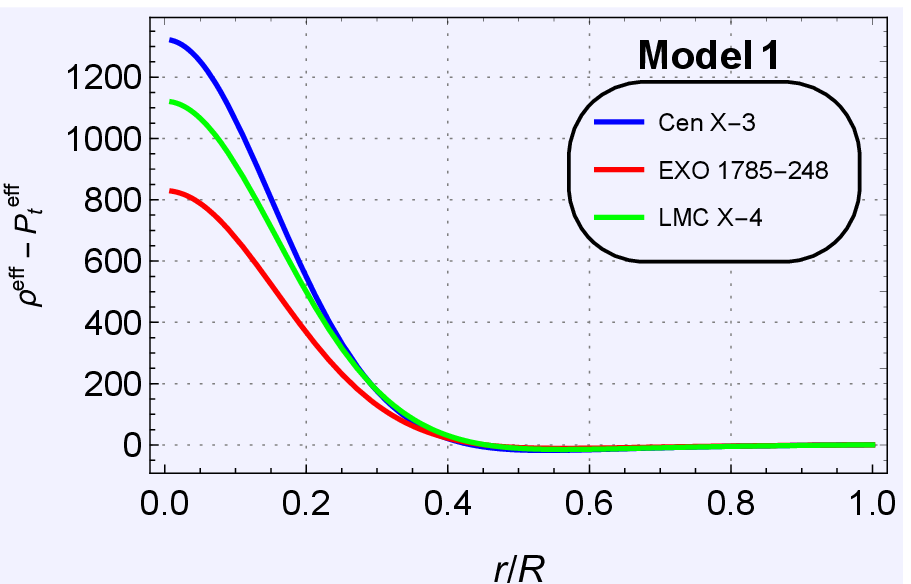,width=0.27\linewidth} &
\epsfig{file=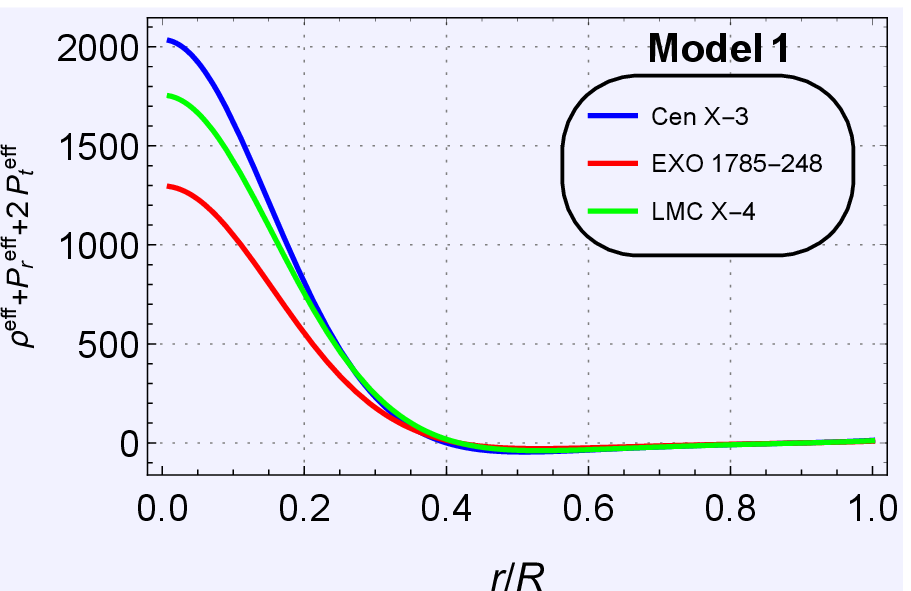,width=0.27\linewidth} &
\end{tabular}
\caption{Plot of energy conditions for stars, $Cen X-3$, $EXO 1785-248$ and $LMC X-4$, under viable $f(\mathcal{G})$ gravity model $1$.}
\label{Fig:8}
\end{figure}
\FloatBarrier

\begin{figure}[h!]
\begin{tabular}{cccc}
\epsfig{file=densitym2.eps,width=0.27\linewidth} &
\epsfig{file=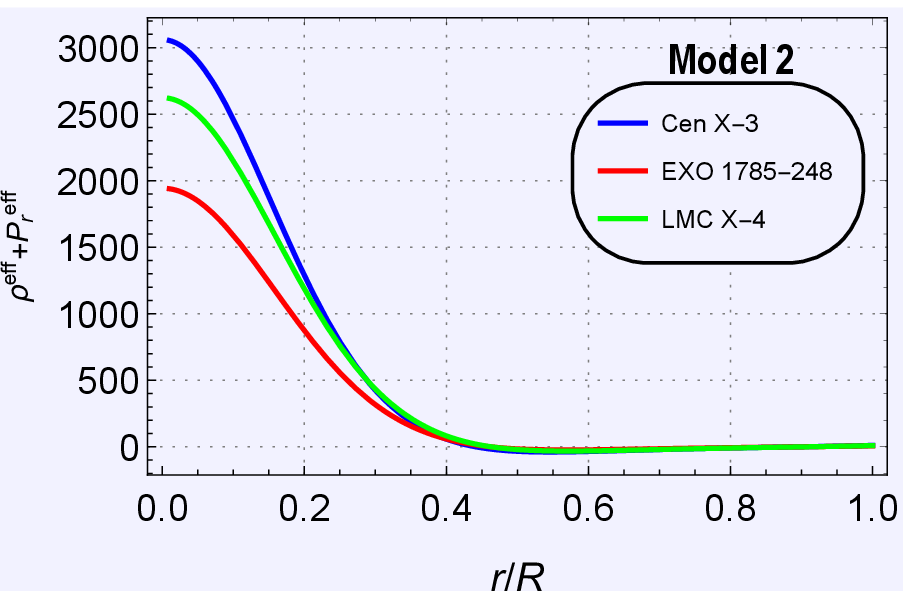,width=0.27\linewidth} &
\epsfig{file=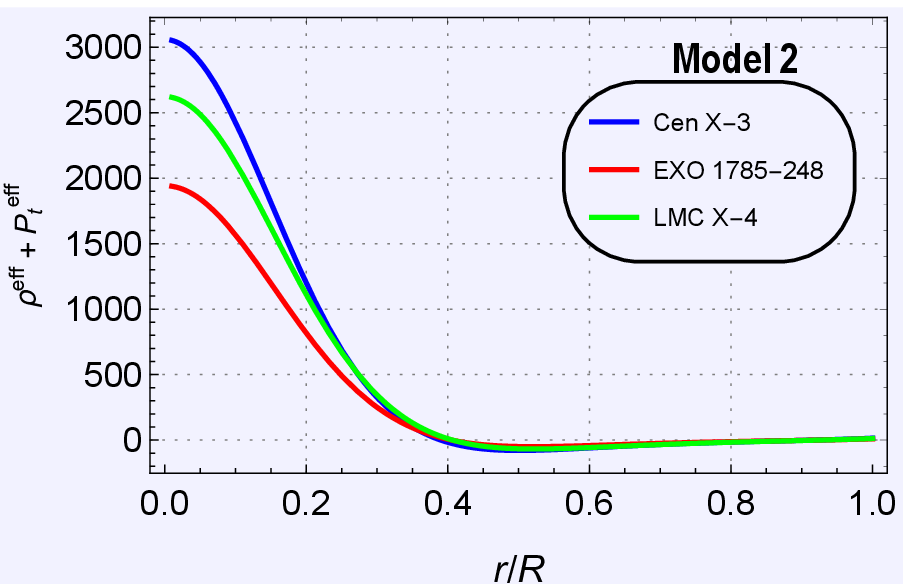,width=0.27\linewidth} &
\end{tabular}
\end{figure}
\FloatBarrier
\begin{figure}[h!]
\begin{tabular}{cccc}
\epsfig{file=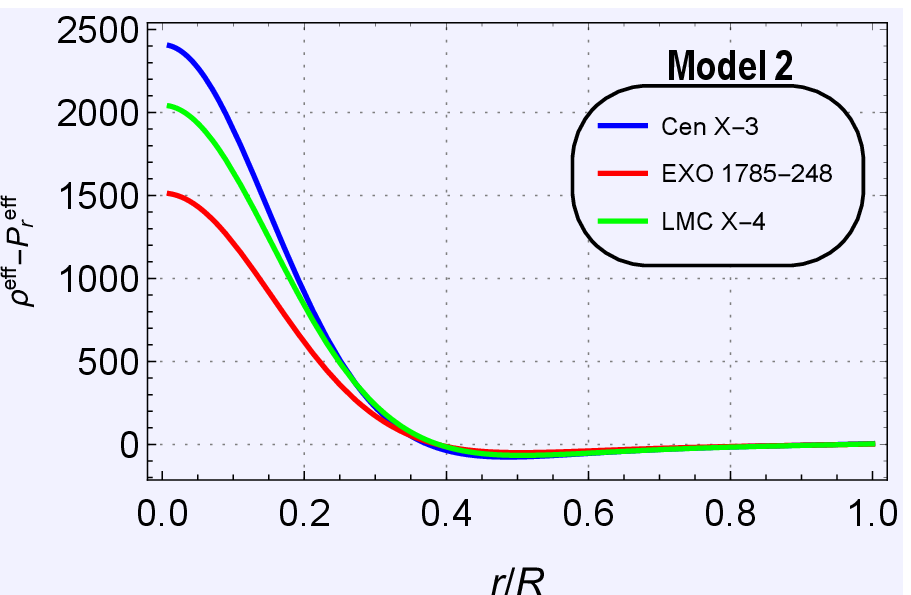,width=0.27\linewidth} &
\epsfig{file=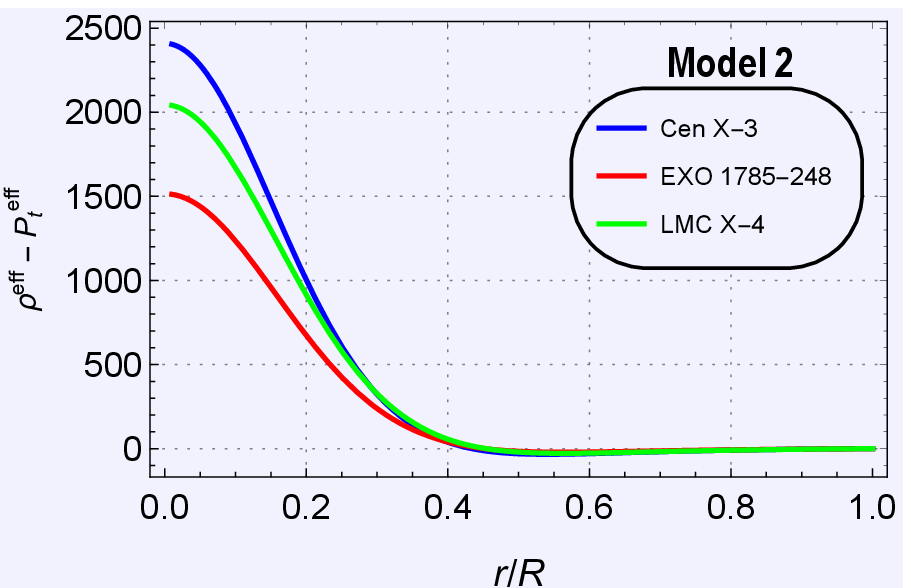,width=0.27\linewidth} &
\epsfig{file=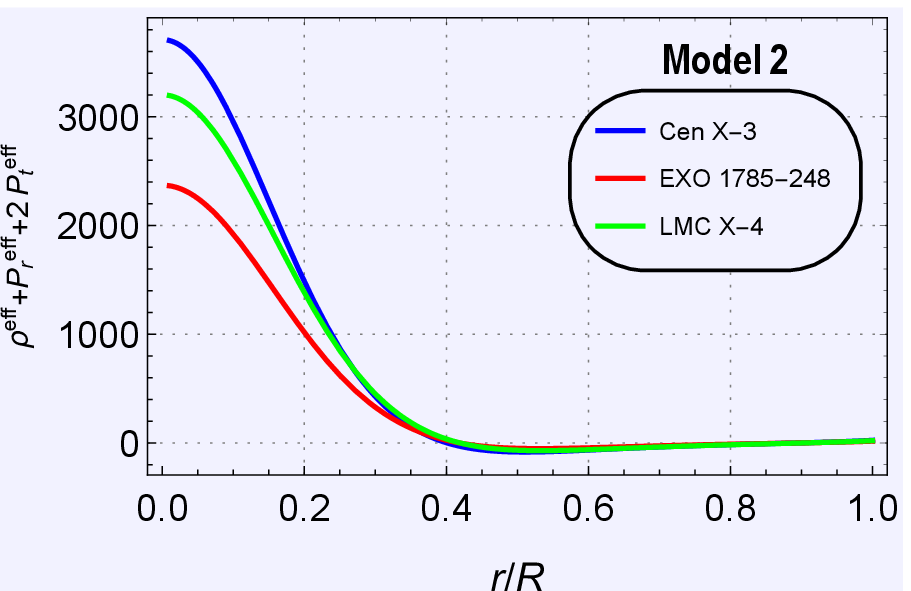,width=0.27\linewidth} &
\end{tabular}
\caption{Plot of energy conditions for stars, $Cen X-3$, $EXO 1785-248$ and $LMC X-4$, under viable $f(\mathcal{G})$ gravity model $2$.}
\label{Fig:9}
\end{figure}
\FloatBarrier
\subsection{ The Modified TOV Equation for $f(\mathcal{G})$ Gravity}
\begin{figure}[h!]
\begin{tabular}{ccccc}
\epsfig{file=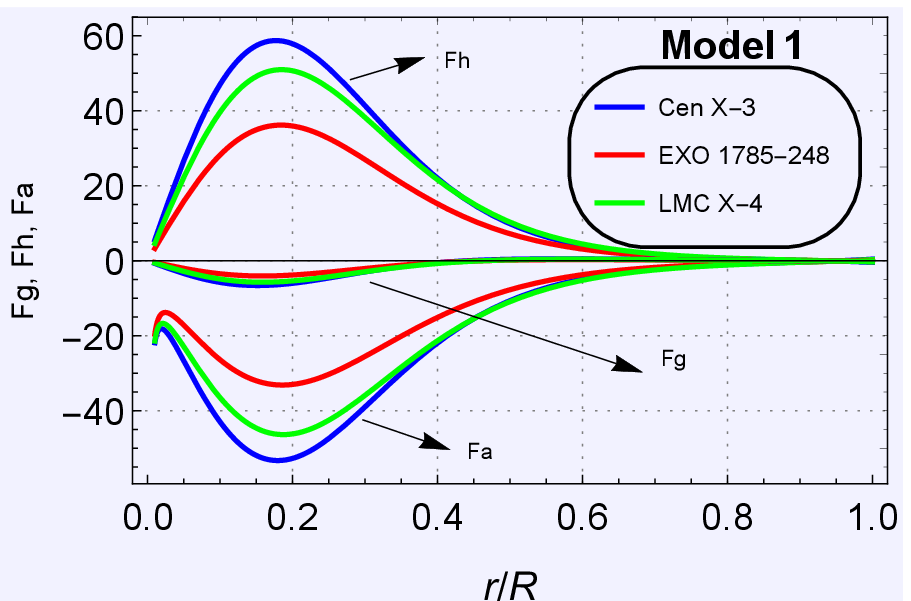,width=0.32\linewidth} &
\epsfig{file=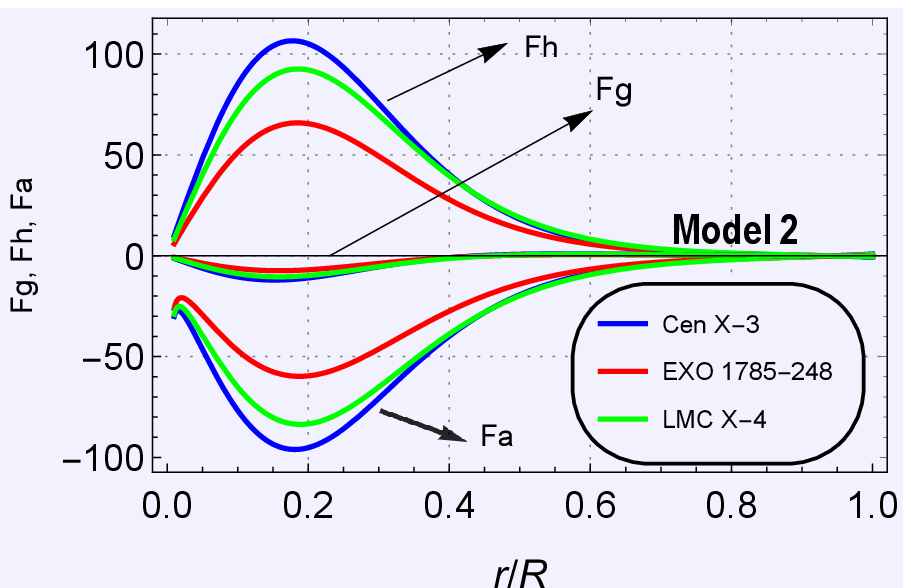,width=0.32\linewidth} &
\end{tabular}
\caption{Behavior of hydrostatic force $(F_h)$, gravitational force $(F_g)$ and anisotropic force $(F_a)$ for stars, $Cen X-3$, $EXO 1785-248$ and $LMC X-4$, under viable $f(\mathcal{G})$ gravity model $1$ on left panel while for model $2$ is on the right panel.}
\label{Fig:10}
\end{figure}

Energy conservation equation of motion for our system is defined by
\begin{equation}\label{23}
\bigtriangledown^\mu T^{eff}_{\mu\nu}=0.
\end{equation}
The modified form of
the generalized TOV equation for
$f(\mathcal{G})$ gravity can be constructed as
\begin{equation}\label{24}
\frac{dp_{r}}{dr}+\frac{\nu~'}{2}(\rho +p_r)+\frac{2}{r}(p_{r}-p_{t})=0.
\end{equation}
The physically acceptable models must be stable under the three forces, viz., gravitational force
$(F_g$), hydrostatic force $(F_h)$ and anisotropic force $(F_a)$ in such a way, that the
sum of the forces becomes zero for the system to be in equilibrium, proposed by
Tolman \cite{Tolman}, and later on Oppenheimer and Volkoff \cite{Oppen} i.e.
\begin{equation}\label{25}
F_g +F_h+F_a=0,
\end{equation}
where $F_g= \frac{\nu~'}{2}(\rho +p_r)$, $F_h= \frac{dp_{r}}{dr}$, $F_a= \frac{2}{r}(p_{r}-p_{t})$.
From the Fig. $\ref{Fig:10}$, it is clear that mutual effect of all forces $F_g, F_h$ and $F_a$ justify the condition of equilibrium for our system.
\FloatBarrier

\subsection{Stability analysis}
\begin{figure}[h!]
\begin{tabular}{cccc}
\epsfig{file=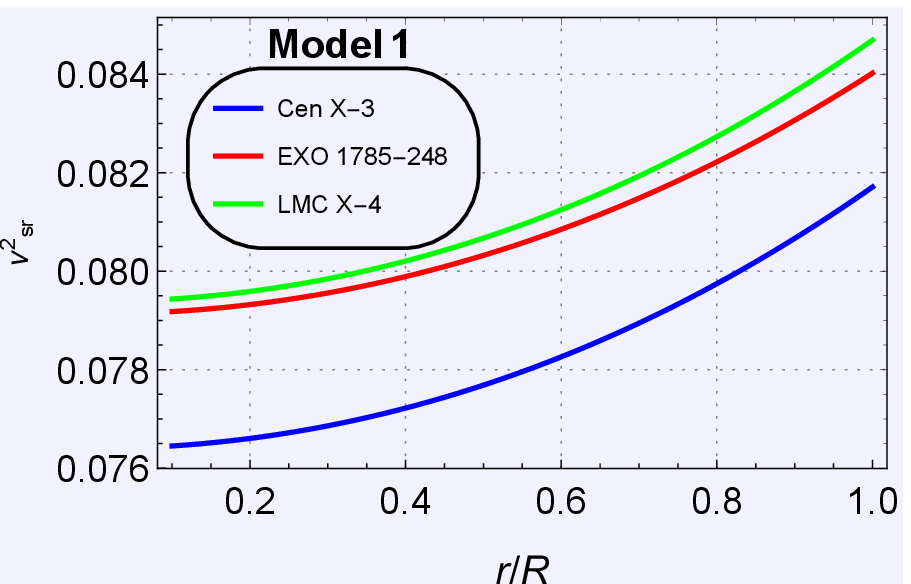,width=0.27\linewidth} &
\epsfig{file=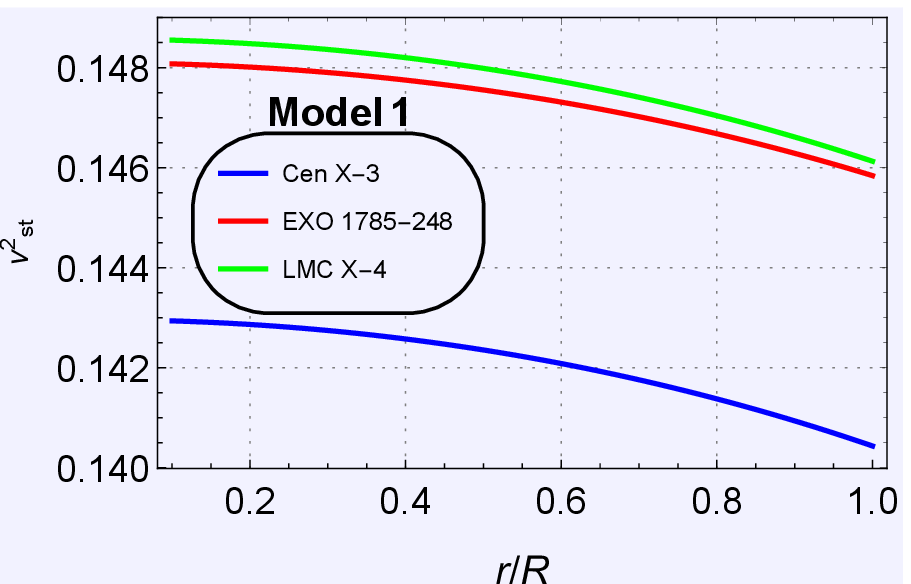,width=0.27\linewidth} &
\epsfig{file=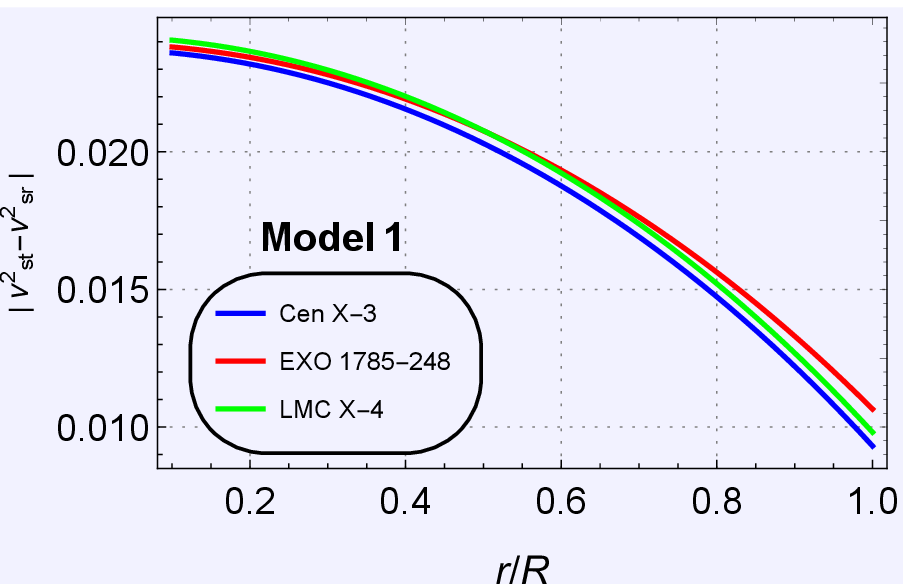,width=0.27\linewidth} &
\end{tabular}
\caption{Variation of $v^2_{sr}$, $v^2_{st}$ and $|v^2_{st}-v^2_{sr}|$
  for stars, $Cen X-3$, $EXO 1785-248$ and $LMC X-4$, under viable $f(\mathcal{G})$ gravity model $1$.}
\label{Fig:11}
\end{figure}

\begin{figure}[h!]
\begin{tabular}{cccc}
\epsfig{file=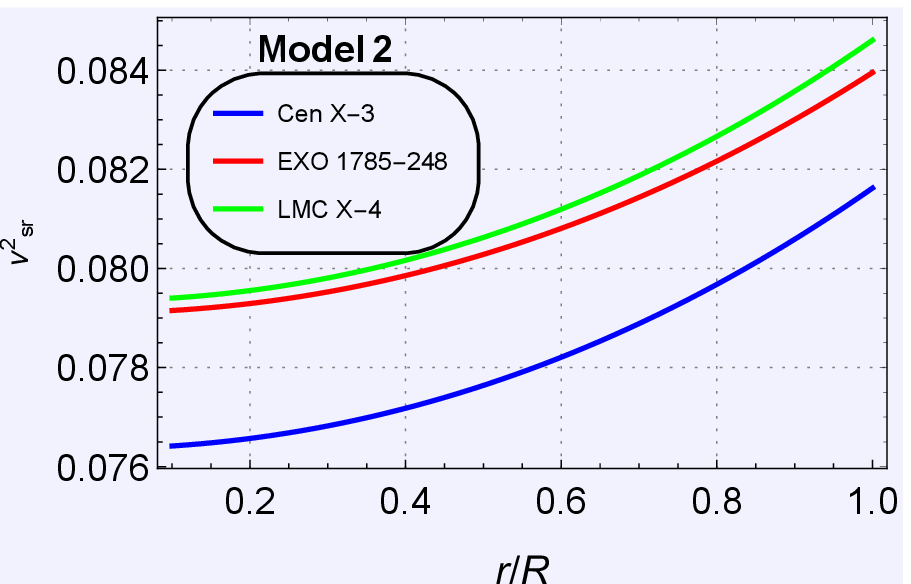,width=0.28\linewidth} &
\epsfig{file=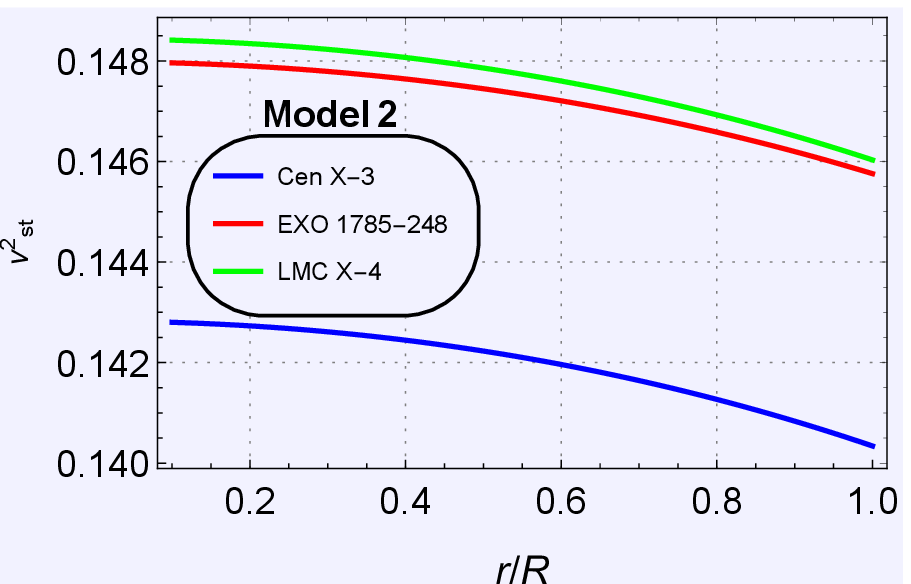,width=0.28\linewidth} &
\epsfig{file=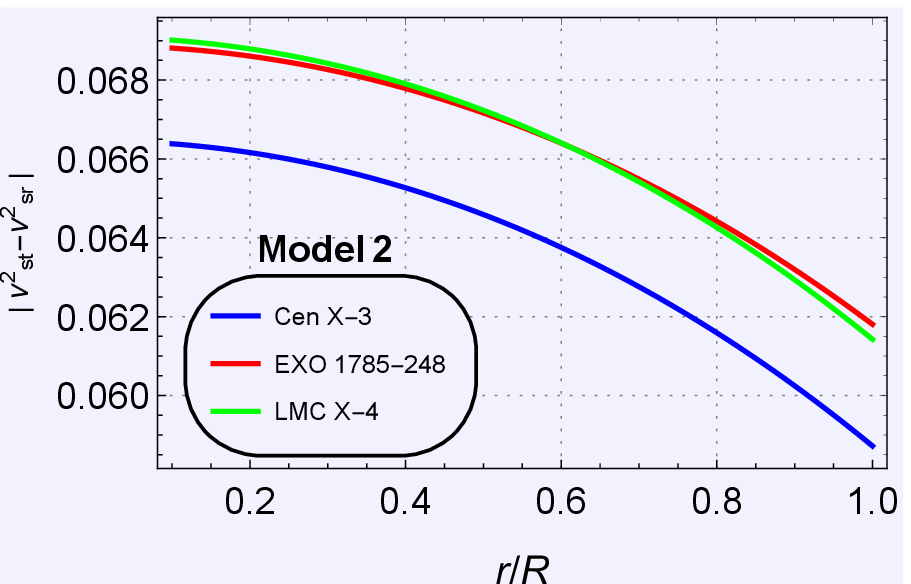,width=0.28\linewidth} &
\end{tabular}
\caption{Variation of $v^2_{sr}$, $v^2_{st}$ and $|v^2_{st}-v^2_{sr}|$
  for stars, $Cen X-3$, $EXO 1785-248$ and $LMC X-4$, under viable $f(\mathcal{G})$ gravity model $2$.}
\label{Fig:12}
\end{figure}

Stellar structure's stability plays an important role in examining
the physical consistent of the models. To study evolution of stellar structure configuration, the role of stability is considered as a very critical and burning issue. A lot of work on stability analysis has been discussed by many researchers. Here we consider Herrera's cracking concept \cite{Her}
to probe the stability of our considered compact stars candidates via technique of radial and
transverse sounds speed symbolized by $v^2_{sr}$ and $v^2_{tr}$. The radial and transverse speed of sounds are defined as
\begin{equation}\label{27}
v^2_{sr}=\frac{dp^{eff}_{r}}{d\rho^{eff}},~~~~~\text{and}~~~~~~ v^2_{st}=\frac{dp^{eff}_{t}}{d\rho}.
\end{equation}
To preserve causality condition, the radial and transverse sound speed
must lie in interval $[0, 1]$ i.e. $0\leq v^2_{sr}\leq1$ and $0\leq v^2_{st}\leq1$, everywhere inside the star, for a physically
stable stellar structure. Herrera and collaborators \cite{Her}-\cite{Prisco}
constructed new notion of the cracking concept to explore the potentially stable/unstable configurations
of stellar structures. One can easily assess that potentially stable/unstable regions within  matter configurations are determined by the difference of sound propagation. The region in which specifically the components of radial speed sound is greater than the components of transverse sound speed
is known as potentially stable region i.e. $0\leq |v^2_{st}-v^2_{sr}|\leq1$, while for unstable region this inequality doesn't hold.
The evolution of radial and transversal speed of sounds for compact star candidates $Cen ~X-3$, $EXO ~1785-248$ and $LMC~ X-4$ can easily seen from the Figs. $\ref{Fig:11}$ and $\ref{Fig:12}$ and it is noted that matter configuration relation is stable, as discussed.
\FloatBarrier
\FloatBarrier
\subsection{Mass-Radius Relationship}
In this section, we scrutinize the mass of compact stars depending upon the radial function $r$, given as
\begin{equation}\label{28}
M^{eff}= \int_{0} ^{R} 4\pi \rho^{eff} r^2 dr.
\end{equation}
The behavior of mass function in Fig. $\ref{Fig:13}$ clearly shows that the mass of compact star is directly proportional to the radius which depicts that mass is regular at core i.e. $M^{eff}\rightarrow0$ as $r\rightarrow0$. Here we can see from the graph that maximum mass is obtained at $r=R$. Furthermore the mass-radius relation in the framework of $f(\mathcal{G})$ gravity is also compatible, while studying the neutron stars \cite{Ast2}.
Moreover, Buchdahl \cite{Buchdahl} found a limit for the mass
to radius ratio for the static spherically symmetric model with anisotropic perfect fluid case, should be bounded like $\frac{2M}{R}<\frac{8}{9}$.
\subsection {Compactification Factor and Redshift Analysis}
The compactification factor $\mu(r)$ can be expressed by  mass to the radius ratio, and defined as
\begin{equation}\label{29}
\mu(r)=\frac{M^{eff}}{r} =\frac{1}{r}\int_{0} ^{R} 4\pi \rho^{eff} r^2 dr.
\end{equation}
\\Furthermore, surface redshift can be determined by the following form
\begin{equation}\label{30}
z_{s}=\frac{1}{\sqrt{1-2\mu}}-1.
\end{equation}

\begin{figure}[h!]
\begin{tabular}{cccc}
\epsfig{file=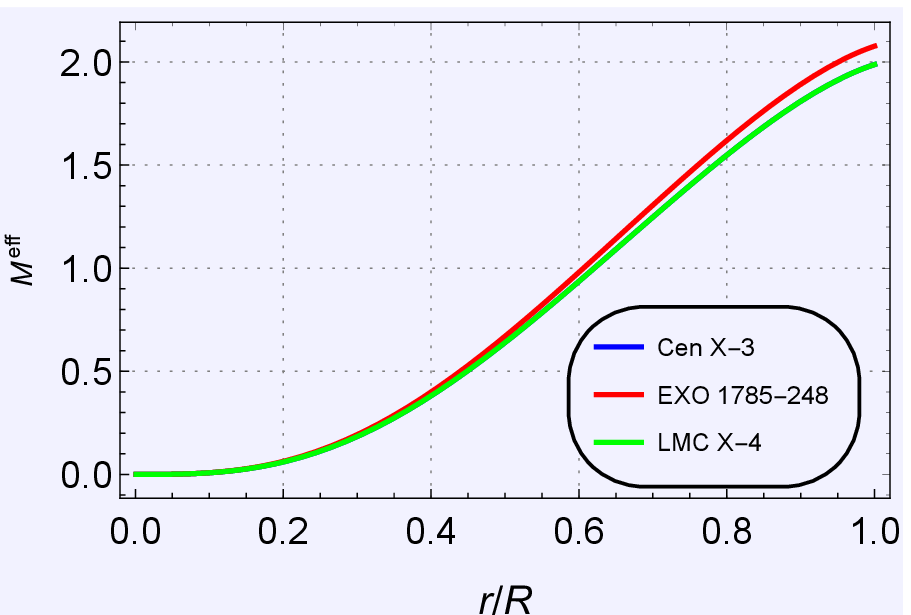,width=0.29\linewidth} &
\epsfig{file=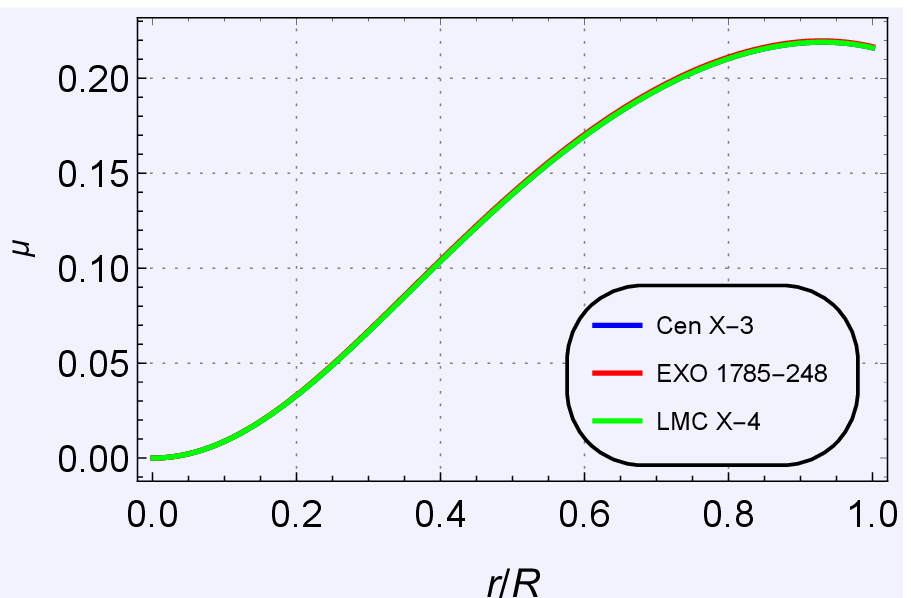,width=0.29\linewidth} &
\epsfig{file=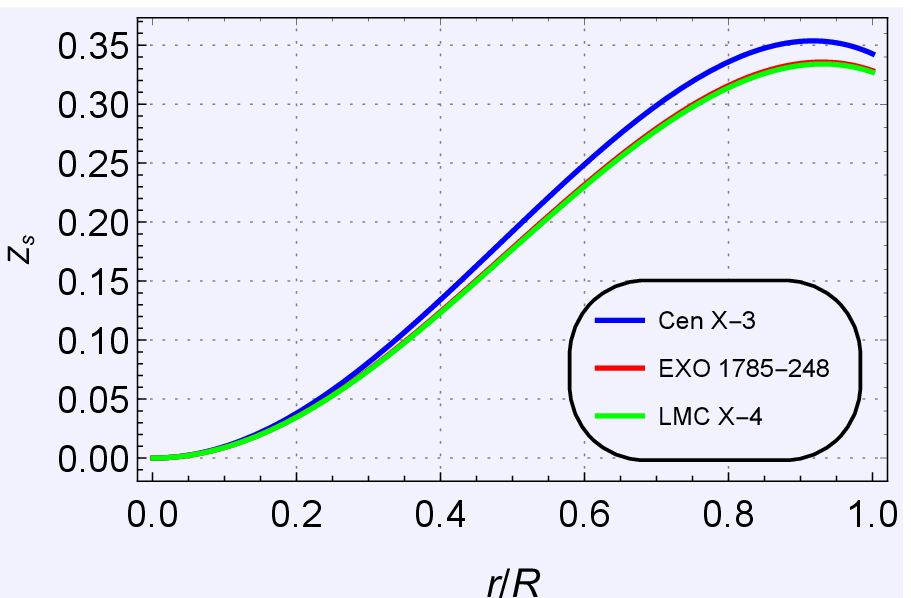,width=0.29\linewidth}  &
\end{tabular}
\caption{Variation of the mass (left panel), compactness factor (middle panel) and redshift (right panel) for $Cen X-3$, $EXO 1785-248$ and $LMC X-4$ compact stars.}
\label{Fig:13}
\end{figure}
\FloatBarrier

The strong physical interaction between particles inside the star
and its equation of state can be described by surface redshift.
The variation of surface redshift and compactness factor for suggested compact stars with
respect to fractional radial coordinate is shown in Fig. $\ref{Fig:13}$.
It increases towards the boundary surface
of compact stars but vanishes
at the center. In our case, all compact stars satisfy Buchdahl condition and the allowed maximum value for surface redshift is $z_{s}\leq5.211$
\cite{B.V}.

\subsection{Adiabatic Index Analysis}
The stiffness of equation of state can be described by the term adiabatic index, for a given energy density, and it also illustrates the stability of the both relativistic and non-relativistic compact stars. The concept of the dynamical stability against
infinitesimal radial adiabatic perturbation of the stellar system has been developed by
Chandrasekhar (as a pioneer) \cite{Chand} and later on this idea has been successfully tested by many authors \cite{Heint}-\cite{Bombaci} for both the isotropic and anisotropic stellar objects.
In their work it is estimated that for a dynamically stable stellar objects the adiabatic index must be
greater than $\frac{4}{3}$ in all internal points. The notation of the
adiabatic index corresponding to radial
and transverse pressure for anisotropic fluid is defined as
\begin{equation}\label{31}
\Gamma_r=\frac{\rho^{eff}+p^{eff}_{r}}{p^{eff}_r}(\frac{dp^{eff}_{r}}{d\rho^{eff}})= \frac{\rho^{eff}+p^{eff}_{r}}{p^{eff}_r}v^2_{sr}.
\end{equation}
\begin{equation}\label{31}
\Gamma_t=\frac{\rho^{eff}+p^{eff}_{t}}{p^{eff}_t}(\frac{dp^{eff}_{t}}{d\rho^{eff}})= \frac{\rho^{eff}+p^{eff}_{t}}{p^{eff}_t}v^2_{st}.
\end{equation}

\begin{figure}[h!]
\begin{tabular}{cccc}
\epsfig{file=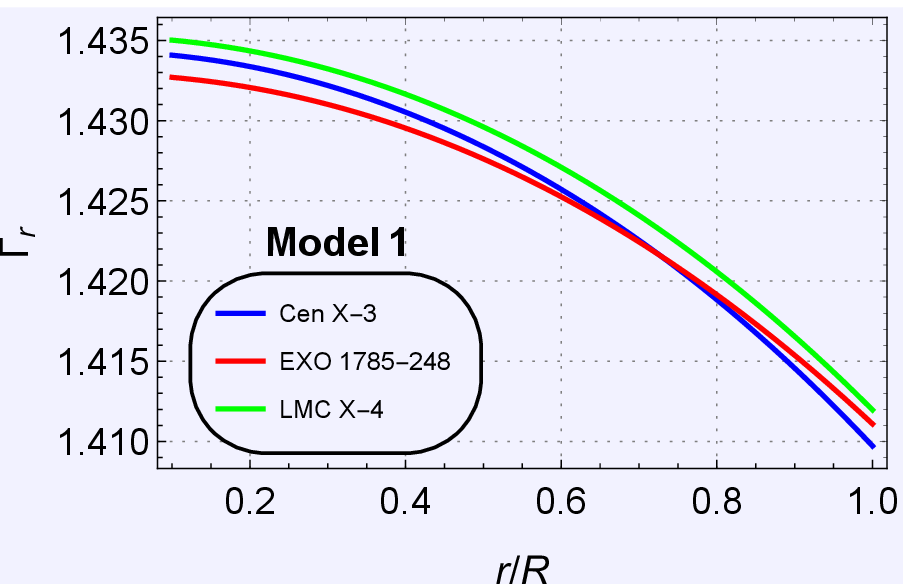,width=0.33\linewidth} &
\epsfig{file=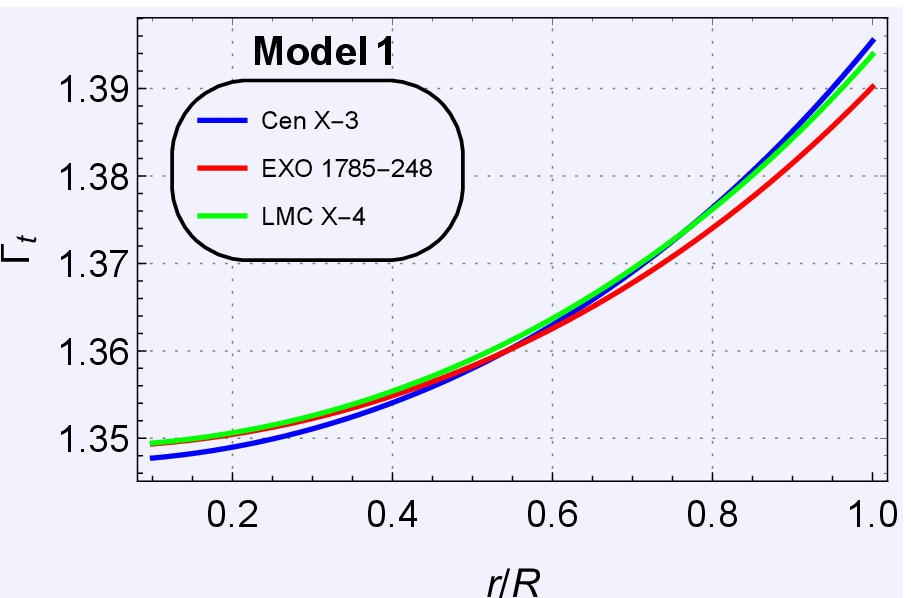,width=0.33\linewidth} &
\end{tabular}
\caption{Variation of adiabatic index for stars, $Cen X-3$, $EXO 1785-248$ and $LMC X-4$, under viable $f(\mathcal{G})$ gravity model $1$.}
\label{Fig:14}
\end{figure}
\FloatBarrier
\begin{figure}[h!]
\begin{tabular}{cccc}
\epsfig{file=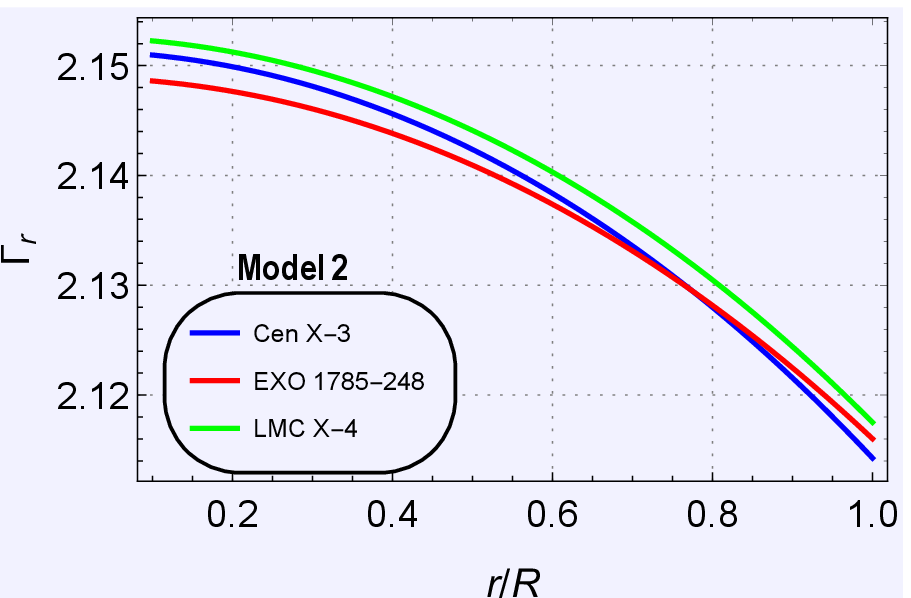,width=0.33\linewidth} &
\epsfig{file=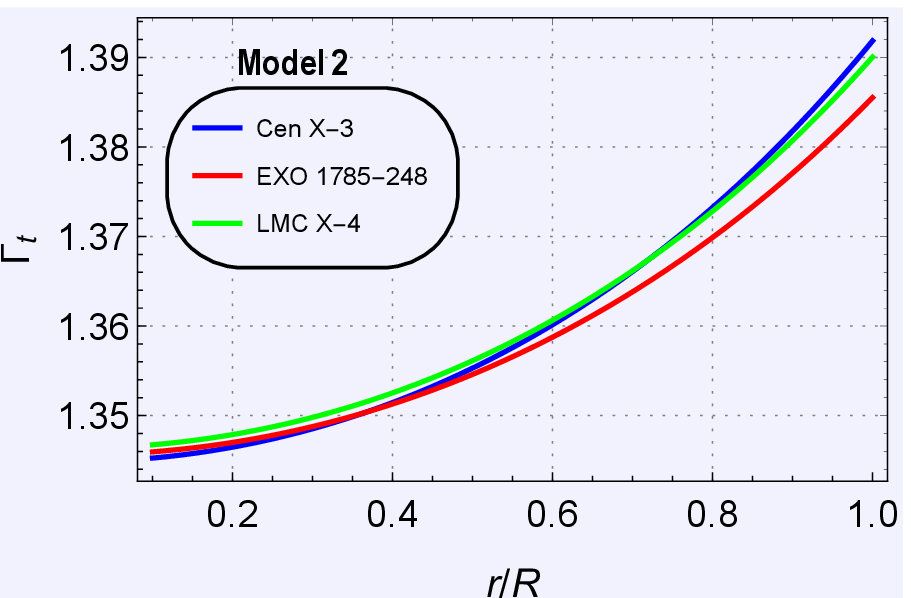,width=0.33\linewidth} &
\end{tabular}
\caption{Variation of adiabatic index for stars, $Cen X-3$, $EXO 1785-248$ and $LMC X-4$, under viable $f(\mathcal{G})$ gravity model $2$.}
\label{Fig:15}
\end{figure}
\FloatBarrier
The behavior of adiabatic index is shown in Figs. $\ref{Fig:14}$ and $\ref{Fig:15}$. From the graph it is clear that the value of adiabatic indices is greater than $\frac{4}{3}$, which confirms the stability of our proposed models.
\subsection{The Measurement of Anisotropy}
\begin{figure}[h!]
\begin{tabular}{cccc}
\epsfig{file=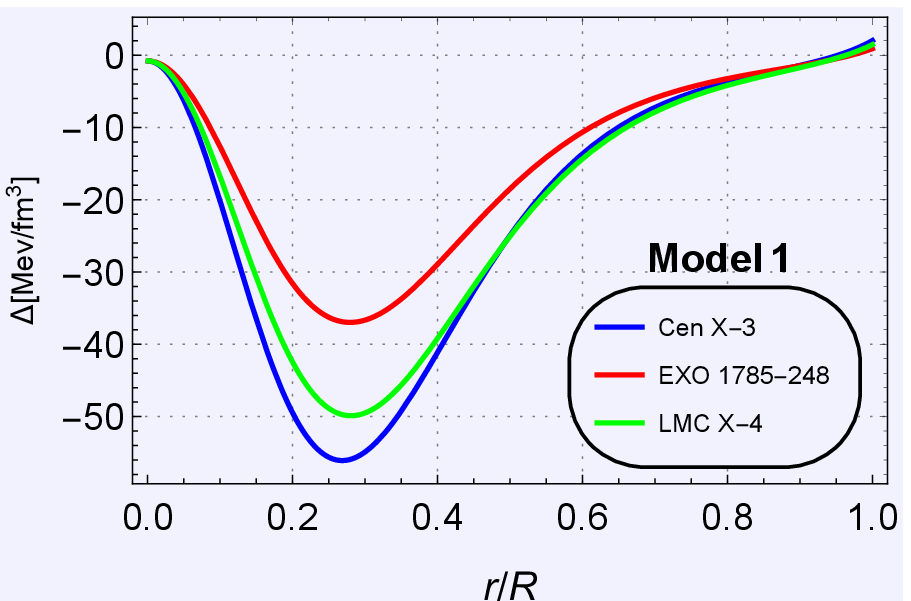,width=0.3\linewidth} &
\epsfig{file=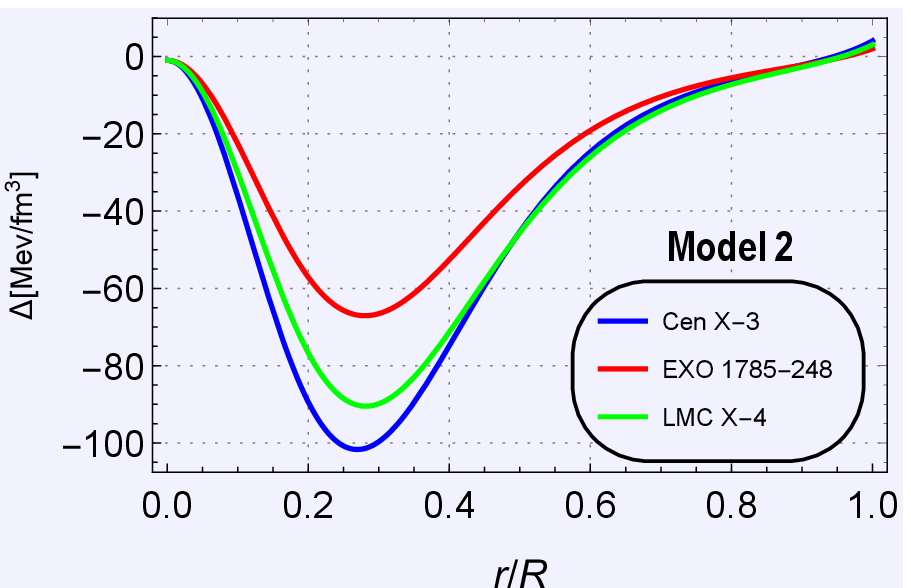,width=0.3\linewidth} &
\end{tabular}
\caption{Variation of the anisotropy factor under viable $f(\mathcal{G})$ gravity model $1$ is on the left panel, while for model $2$ is on the right panel.}
\label{Fig:16}
\end{figure}

In case of compact star modeling the interior structure of relativistic stellar objects can be illustrated by the term anisotropy, expressed as
\begin{equation}\label{32}
\triangle=\frac{2}{r}(p^{eff}_{t}-p^{eff}_{r}),
\end{equation}
which yields the information as regards the anisotropic behavior of the model.
We check the anisotropy behavior graphically with the help of observationally data of the considered stars, which is presented in Table $1$. If $p^{eff}_{t}>p^{eff}_{r}$ then this depicts that anisotropic pressure is directed outward and this leads to $\triangle>0$, while if $p^{eff}_t<p^{eff}_r$ the anisotropy turns negative i.e. $\triangle<0$,  and this identifying that anisotropic being drawn inward.
The graphical analysis of anisotropic measurement corresponding to fractional radial coordinate
$r/R$ shows decreasing behavior for the considered stars, suggesting that $p^{eff}_t < p^{eff}_r$, as shown in Fig. $\ref{Fig:16}$.
\FloatBarrier
\section{Concluding Remarks}
In order to detect the equitable model for
realistic geometry of internal compact stellar structures not
only in GR but also in modified $f(\mathcal{G})$ gravity, is considered as an attracting challenge.
Our motivation is to examine the real composition of these compact stars in
their internal cores under the consideration of two different
viable $f(\mathcal{G})$ gravity models.
For this goal, we have certified these models for
three different observed compact stars, labeled as
$Cen~ X-3$, $EXO ~1785-248$ and $LMC ~X-4$ with an anisotropic source matter by considering the Tolman- Kochowicz spacetime \cite{Jasim}, with metric potentials  $\nu=Br^2+2lnC$ and $\lambda=ln(1 +
ar^2+br^4)$, where $a$, $b$, $C$ and $B$ are constant parameters. These arbitrary constant values are constructed by matching the interior of a metric with Schwarzschild's exterior metric. This aspect is very valuable to observe the physical behavior of the compact stars by indicating their radii and masses in terms of the arbitrary constants.
\\The main aim of this study is to deal the compact stars with formulation of analytical
models by considering anisotropic static source configurations in the context of modified $f(\mathcal{G})$ gravity.
The graphical analysis and interpretation of these results exhibit some conspicuous properties related to these
anisotropic compact stars as follows:
\\
\begin{itemize}
\item Geometry of the space time is described by the metric potentials. The evolution of metric potentials
$e^{\nu}$ and $e^{\lambda}$ with respect to the fractional radial coordinate $r/R$ in Fig. $\ref{Fig:1}$ satisfy the condition
$e^{\lambda(r=0)}=1$  and $e^{\nu(r=0)}=C^2$. Both metric potentials have minimum value at the core of star and then increase monotonically away from the center to surface. For a physical viability and stability of the suggested models, metric potentials should be positive, finite and free from the geometrical singularities.
The graphical behavior in Fig. $\ref{Fig:1}$ clearly shows that our metric potentials are consistent and satisfy all the above requirements.

\item The variation of effective energy density, radial and tangential pressure corresponds to the fractional radial coordinate $r/R$
for both models is regular at the center. The graphical behavior depicts that the effective energy density and both pressures are free from the central singularities. It is clear from the Figs. $\ref{Fig:2}$ and $\ref{Fig:3}$ that these features attain maximum value at the center and show continuously decreasing behavior away from the center to the boundary of the star. The numerical values of effective energy density and radial pressure for suggested models at the center of the three compact objects namely $Cen~ X-3$, $EXO ~1785-248$ and $LMC~ X-4$ are given in Table $\ref{tab2}$ and $\ref{tab3}$. These numerical values clearly show that the value of central density is higher than the surface density.
 This fact assures the high compactness of most dense stellar compact objects.
\item It has been observed from the Figs. $\ref{Fig4}$ - $\ref{Fig7}$ that the radial derivatives of effective energy density and anisotropic pressures are negative, and at the center these values vanish. This fact confirms high compactness at core
of the star.

\item It can be noted from the Figs. $\ref{Fig:8}$ and $\ref{Fig:9}$ that all energy bounds for our proposed models are well satisfied which exhibit
the realistic matter content.

\item  To check that whether all forces namely, gravitational force
$(F_g$), hydrostatic force $(F_h)$ and anisotropic force $(F_a)$ are in equilibrium for our models, we studied the TOV equation in modified $f(\mathcal{G})$ gravity frame of reference. Fig. $\ref{Fig:10}$  yields that all the forces are in equilibrium, which endorse
stability of our system.

\item The radial and tangential speeds of sound for compact stars are denoted by $v^2_{sr}$ and $v^2_{st}$. The values of the square of sound speeds  lie within the range $0$ and $1$. From the Figs. $\ref{Fig:11}$ and $\ref{Fig:12}$, it can be easily seen that our models are consistent with the causality condition. Further, our present system is consistent with the Herrera cracking condition i.e. $0\leq |v^2_{st}-v^2_{sr}|\leq1$, which confirms the stability of our system.

\item Fig. $\ref{Fig:13}$ shows that the calculated mass for our suggested models is very close to the standard observational data, which depicts that our mass function is regular at the center of core. Further, the evolution of compactification factor and the behavior of surface redshift with respect to the fractional radial coordinate favor our models, as the values of compactness and redshift factor satisfy the required limits.

\item  The radial $\Gamma_r$ and tangential $\Gamma_t$ adiabatic indices have been constructed in Figs. $\ref{Fig:14}$ and $\ref{Fig:15}$.
For a dynamically stable stellar objects the adiabatic index must be
greater than $\frac{4}{3}$ in all internal points. It is clear from the graphical representation that the value both adiabatic indices are greater than $\frac{4}{3}$ throughout the star, which establish the stable nature of our proposed models.
 \\
\end{itemize}
In the study of compact stellar structures, the role of modified  $f(\mathcal{G})$  gravity is very alluring.
The study of possible existence of compact stars and particle physics with their extremely dense cores has enforced the researchers for more reliable solutions of the modified field equations. As a final comment, in this present study we have favorably showed
singularity free and entirely stable stellar system, which is advisable to express
the anisotropic nature of compact stars, by employing the Tolman- Kuchowicz metric. Here we observed that our proposed models in the framework of modified  $f(\mathcal{G})$  gravity are consistent and stable, as all physical attributes of compact stars follow physically accepted patterns.

\end{document}